\pdfoutput=1

\documentclass[12pt,a4paper]{article}
\usepackage[top=1.2in, left=0.8in, right=0.8in, bottom=1.2in]{geometry}
\usepackage{booktabs} 
\usepackage{textgreek} 
\usepackage[ruled]{algorithm2e} 
\usepackage{algorithmic}		
\usepackage{amsmath} 
\usepackage{multirow}
\usepackage[numbers]{natbib}
\usepackage{graphicx}
\usepackage{float}
\usepackage{subfloat}
\usepackage{subfig}
\usepackage{txfonts}	
\usepackage[T1]{fontenc}

\newcommand{\footremember}[2]{%
    \footnote{#2}
    \newcounter{#1}
    \setcounter{#1}{\value{footnote}}%
}
\newcommand{\footrecall}[1]{%
    \footnotemark[\value{#1}]%
} 

\author{%
   Georgios Fortetsanakis\footremember{trick}{University of Crete and Foundation for Research and Technology-Hellas, Greece.}
   \and Maria Papadopouli \footrecall{trick} \footremember{second trick}{Contact author: Maria Papadopouli (\texttt{mgp@ics.forth.gr)}.} 
  }
\date{}

\begin{document}
\title{Multi-layer Game-Theoretical Analysis of Wireless Markets with Market Segmentation}
\maketitle
\begin{footnotesize}
\noindent
{\bf Abstract}
New wireless access markets have emerged that are larger, more heterogeneous and diverse. Modelling such markets can be challenging due to the interplay of various business- and network-related aspects as well as the interdependencies among different entities (e.g., customers, providers). Existing models of wireless markets are either microscopic, focusing on a specific technical aspect (e.g., protocol, network topology, technology) at a fine scale or macroscopic modelling wireless markets at a large-scale, e.g., considering homogeneous user populations. In contrast to these approaches, this work develops a multi-layer game-theoretical framework, which allows providers to model users at multiple levels of detail by considering a different number of user sub-populations. It also models the mobility pattern, traffic demand, and  networks of providers. A population game using Logit dynamics models the user selection of the appropriate dataplan and provider, capturing the diversity in customer profile and relaxing the assumption about the user rationality. It {\em analytically} computes the equilibriums of users and providers and numerically evaluates the performance of the market as a function of the traffic demand, the number of available dataplans, and the knowledge about customer population. 
Significant benefits in revenue can be achieved by a provider when it integrates more detailed information about the user population. The number of disconnected users also decreases. Moreover the availability of several dataplans further enhances the gains. The stronger the provider, the more prominent the benefits. However the benefits diminish when all the providers model the customer population at the same degree of detail due to an increased competition. The analysis highlights the development of different strategies of the providers depending on their capacity, level of knowledge about the customer population, and traffic conditions. It illustrates how a provider changes its strategy under different conditions, focusing potentially on different customer segments and also the pressure introduced by specific customer types.
\end{footnotesize}

\section{Introduction}\label{sec:introduction}

The wireless markets have experienced drastic changes and expansion, becoming more diverse (e.g., in terms of customer profiles, services, providers), more complex and dynamic. 
The virtualization, increased mobile traffic, cloud infrastructures,  and OTT applications have been changing the landscape.
The telecom industry is at a crossroads, facing competitive challenges, such as the decline of legacy services 
(e.g., voice and SMS have been rapidly supplanted by OTT applications), increased WiFi access, OTT competition, and saturation, that further enhances the competitive pricing pressure \cite{accenture}.
Pricing can be used to regulate the user behaviour, manage the network resources, while optimizing the Quality of Service (QoS) and revenue, especially as the demand may exceed the capacity. 
Therefore the design of appropriate pricing plans is important: high prices may result in high churn, while low prices may substantially increase the number of subscribers, 
which may result in congestion and customer dissatisfaction.  
A good balance between user requirements and profit can be achieved by the identification of optimal plans. 
For that, various economic and technical aspects that affect the customer and provider decision making need to be modelled. 
Providers have applied various data-mining algorithms to identify the different market segments (e.g., \cite{Sche10, Han12, Ye13, Kian06}). 
How does the knowledge about the different market segments impact the churn and the profit? How do providers develop their strategies and how do they change them to face the traffic increase and improve their profit? How does a weak provider ``survive'' in the presence of stronger providers? How do the different market segments influence the market?
Such questions can be answered using game-theory.

In general the game-theoretical modelling of wireless markets is performed either at a microscopic or at a macroscopic level: Microscopic approaches usually focus on specific 
technical aspects (e.g., protocol, network topology, technology) on a short spatial and temporal scale, often considering networks with a small number of base stations (BSs) 
and markets with a single provider \cite{Gaji14, Dape11, Yang11, Gao0411, Yu0510, Gao14, Zhuo14}. The computational and scalability issues when analysing large (e.g., nation-wide) markets become prominent.
On the other hand, macroscopic approaches model large-scale markets using homogeneous populations \cite{Jia09, Niya0710, Niya09, Niyat08, Jia0508} 
and a simplified version of the network infrastructure \cite{zhan17,zhan18} to make the analysis tractable. Such macroscopic approaches alleviate the computational issues but at the expense of accuracy, since they do not consider the diversity in the customer population.
Unlike these approaches, this work develops a {\em multi-layer} two-stage game-theoretical framework that allows providers to model users at 
multiple levels of detail based on their profile. Our objective is twofold: to answer the aforementioned questions and address the computational-accuracy trade-off. 

%

The competition of providers is modelled as a normal-form game, in which providers strategically select their prices to optimize  their revenue.
The framework models in detail the network of several providers, each offering multiple pricing plans to their customers. 
The plans are data-oriented (with prices based on data caps) and the focus is on wireless data traffic, as in most telecom markets currently. 
The framework captures important aspects of the customer behaviour, 
including their preferences and loyalty, traffic demand, mobility pattern (e.g., session arrival process, handovers), and QoS metrics. It identifies market segments (i.e., user sub-populations) based on their willingness-to-pay, 
datarate, and traffic demand, and model their decision making separately. 
Examples of such choice strategies consumers may use, especially when price is better known than quality, are the best value, price-seeking, and price aversion \cite{tell90}.
A population game models the user decision-making process. Via the Logit dynamics, a user decides to become subscriber of a certain provider or remain disconnected.

Based on this modelling framework, we analysed the benefits of integrating information about the market segments (e.g., consumer profiles) 
in different levels of detail (i.e., multi-layer aspect). The evaluation focuses on the revenue of providers and percentage of disconnected users, as a function of the level of detail, number of dataplans, and traffic demand. 
Significant benefits in revenue can be achieved by a provider when it integrates more detailed information about the user population. The number of disconnected users also decreases. 
Moreover the availability of several dataplans further enhances the gains.  The stronger the provider, the more prominent the benefits. However the benefits diminish when all the providers model the population at the same degree of detail, due to an increased competition.
The analysis highlights the development of different strategies of the providers depending on their capacity, level of knowledge about the customer population, and traffic conditions. It illustrates how a provider changes its strategy under different conditions, focusing potentially on different customer segments and also the pressure introduced by specific customer types. It also shows how the weak provider reacts and ``survives" the competition. With a larger number of dataplans, providers can charge the different customer sub-populations more efficiently, achieving higher revenue.

This paper extends our earlier work and the state-of-the-art in the following ways: 
(1) It enables providers to model users at {\em different levels of detail} and {\em analytically} computes the equilibriums of users and providers. 
(2) It employs a detailed model of the network. 
(3) It allows providers to offer several dataplans depending on user traffic demand and models the user-decision making, capturing important aspects of customer behaviour.
(4) It relaxes the assumption about the user rationality in selecting providers and dataplans. 
The irrationality is often attributed to the subjective assessment on the benefits and risks (as indicated by behavior economics). 

The paper is structured as follows: Section \ref{sec:Modeling_framework} presents our modeling framework, focusing on the queuing network of providers, user service selection, and the competition of providers. Section \ref{sec:Perf_evaluation} evaluates the performance of a wireless market when providers model users at different levels of detail. Section \ref{sec:Related_work} discusses the related work. Finally, Section \ref{sec:Conclusion} presents our conclusions and future work plan.

\section{Modeling framework} \label{sec:Modeling_framework}
Economic and technical aspects affect the decision-making process of customers and of providers.
Our modelling framework consists of two layers, the technological layer and the economic one. The technological layer models the wireless networks of providers as queueing networks and the user traffic demand with appropriate stochastic processes. It also estimates the QoS of providers based on the average and variance of data rate. 
The economic layer models the interaction among a set of providers and a heterogeneous user population. Each provider offers different dataplans to users that produce different traffic demand and selects their prices aiming to maximize its revenue, while each user selects a provider considering the offered prices and quality of service (QoS) guarantees. The user population is divided into groups, each having different characteristics and preferences. Throughout this paper, we use the terms users and customers interchangeably  (similarly for the terms group and segment).
Given the growth of LTE networks, the rapid increase of the WiFi mobile traffic (expected to consitute more than 46 percent of mobile data\cite{cisco17}), the evolution of the IP networks, and the decline of the voice and sms, we focus here on data (downlink) transmission and monthly dataplans.

\begin{figure}[t!]
\centering
\includegraphics[width=4in]{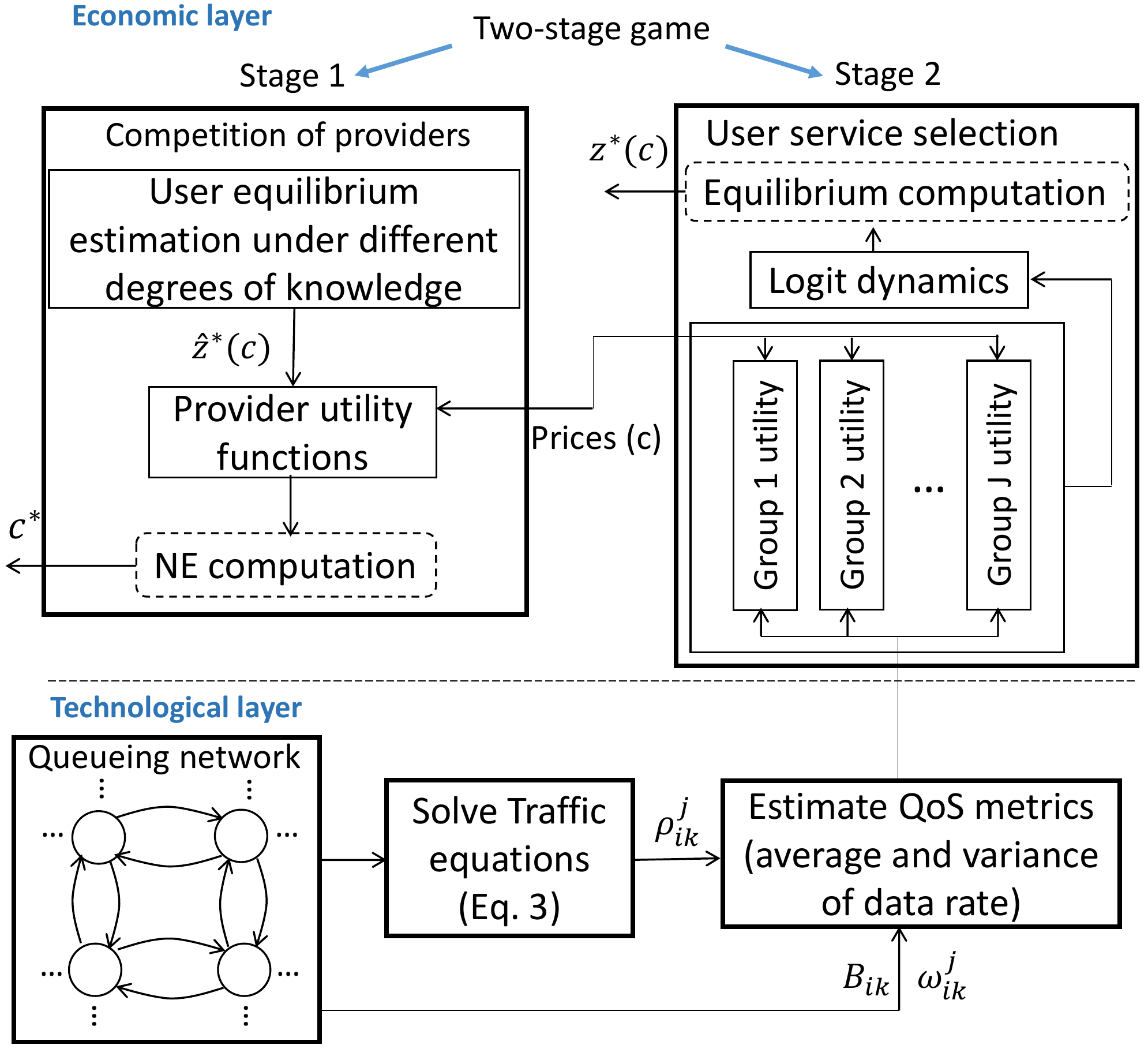}
\caption{Overview of the main components of the framework: network of providers, market segmentation, traffic, customer selection, and pricing strategy of providers.} 
\label{fig:models}
\end{figure}

A {\em two-stage game} defines the interaction of users and providers. The first stage instantiates the {\em competition of providers} and the second one the {\em user-decision making}. A {\em population game} models the user decisions: members of each user group could either select to become subscribers of a certain provider or remain disconnected. The decisions of these users are modeled by the Logit dynamics, a system of ordinary differential equations. They are based on a utility function that depends on the price and QoS and a noise parameter that defines how much trust users place on this utility function, capturing the user ``irrationality'' and ``stickiness'' to a provider. On the other hand, the competition of providers is modeled as a {\em normal-form game} in which providers strategically select their prices to optimize their revenue. The utility functions of providers depend on the offered prices and the equilibrium of users (Fig. \ref{fig:models}). Our framework models a wireless access market of $I$ providers and a user population of $J$ groups. Each provider has deployed a network of wireless BSs and offers long-term subscriptions, which are {\em best-effort data services}. The following subsections describe the components of our modeling framework in more detail.
\begin{table}[!t]
\centering
\captionsetup{justification=centering}
\caption{\textbf{Queueing-theoretical parameters for users of group j when connected at the network of provider i}}
\begin{tabular}{|c|c|}
\hline \textbf{Parameter} & \textbf{Description}\\
\hline \centering $K_i$&Number of BSs of the provider $i$\\
\hline \centering $\lambda_j$&Total session generation rate\\
\hline \centering $z_{ji} (z_{j0})$&Ratio of subscribers (disconnected users)\\
\hline \centering $\omega_{ik}^j$&Steady-state probability for a user of
group $j$ to be located within the coverage of BS $k$\\
\hline \centering $v_{ik}$&Departure rate from BS $k$ due to handover\\
\hline \centering $\mu_{ik}$&Session service rate at BS $k$\\
\hline \centering $d_{ik}$&Total departure rate from BS $k$ ($d_{ik}=v_{ik}+\mu_{ik}$)\\
\hline \centering $p_{i,m,k}^{(j)*}$&Conditional prob. of handover from BS $m$ to BS $k$ given that a handover occurs\\
\hline \centering $p_{i,m,k}^{(j)}$&Unconditional prob. of handover from BS $m$ to BS $k$ $\left(p_{i,m,k}^{(j)}=v_{im}p_{i,m,k}^{(j)*}/d_{im}\right)$\\
\hline \centering $\gamma_{ik}^j$&Total session arrival rate at BS $k$\\
\hline \centering $a_{ik}^j$&Arrival rate of new sessions at BS $k$\\
\hline \centering $\rho_{ik}^j$&Traffic intensity at BS $k$\\
\hline \centering $n_{i}$&Vector indicating the number of users at each BS\\
\hline \centering $Q_i(n_i)$&Stationary distribution of number of users at BSs\\
\hline \centering $B_{ik}$&Bandwidth at BS $k$\\
\hline \centering $R_{ji}(z)$&Average data rate\\
\hline \centering $V_{ji}(z)$&Variance of data rate\\
\hline
\end{tabular}
\label{tbl:technology_parameters}
\end{table}

\subsection{The queueing networks of providers} \label{subsec:queueing_networks}

Each provider (e.g., provider $i$) has deployed a number of BSs ($K_i$) covering a geographical region (e.g., a city). We assume that in all BSs, the available bandwidth is shared equally among connected users, i.e., {\em processor-sharing discipline}. \footnote{Various scheduling algorithms that perform a long-term proportional fair channel allocation have been proposed in the context of LTE networks \cite{Capo13, Kawser12}Similarly, the IEEE 802.11 achieves a long-term fair bandwidth allocation.}. Users generate requests to connect to a base station (BS) to start a session. During a session, a user transmits and receives data via that BS. The session generation of a group of users $j$ follows a Poisson process with a total rate of $\lambda_j$. This rate is allocated across providers according to the probability vector $z_j = (z_{j0}, z_{j1}, ...,z_{jI})$. The ratio of members of the group $j$ that select the provider $i$ is indicated by $z_{ji}$, while $z_{j0}$ indicates the ratio of members of that group that select the disconnection. The vector $z = (z_1,...,z_J)$ corresponds to the probability vector of each user group and shows how the entire user population is divided among the available providers and disconnection state.

\begin{figure*}[t!]
\centering
\subfloat[]{\label{fig:traffic_equations}\includegraphics[width=2.7in]{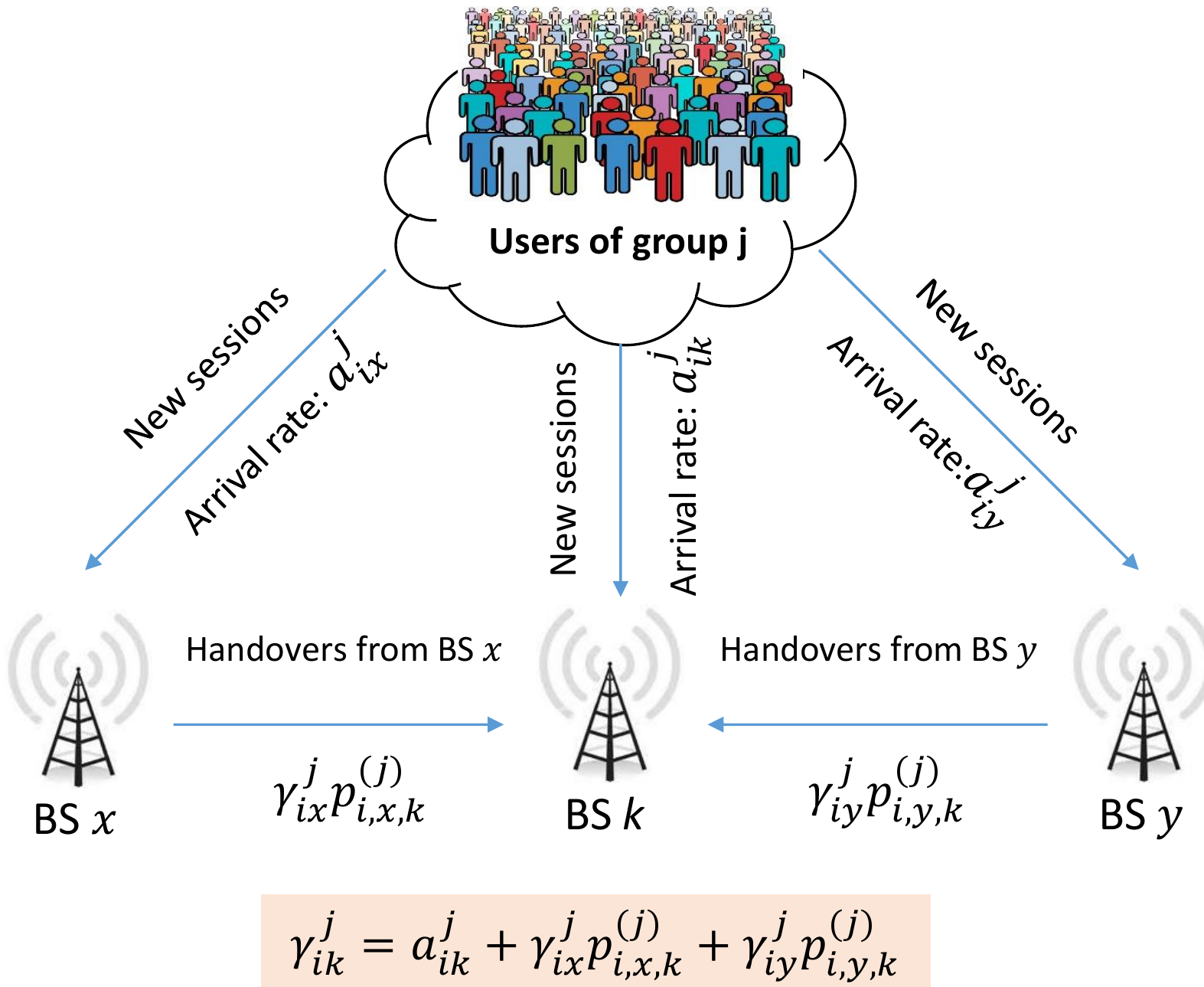}}
\subfloat[]{\label{fig:Local_balance_equations}\includegraphics[width=3.7in]{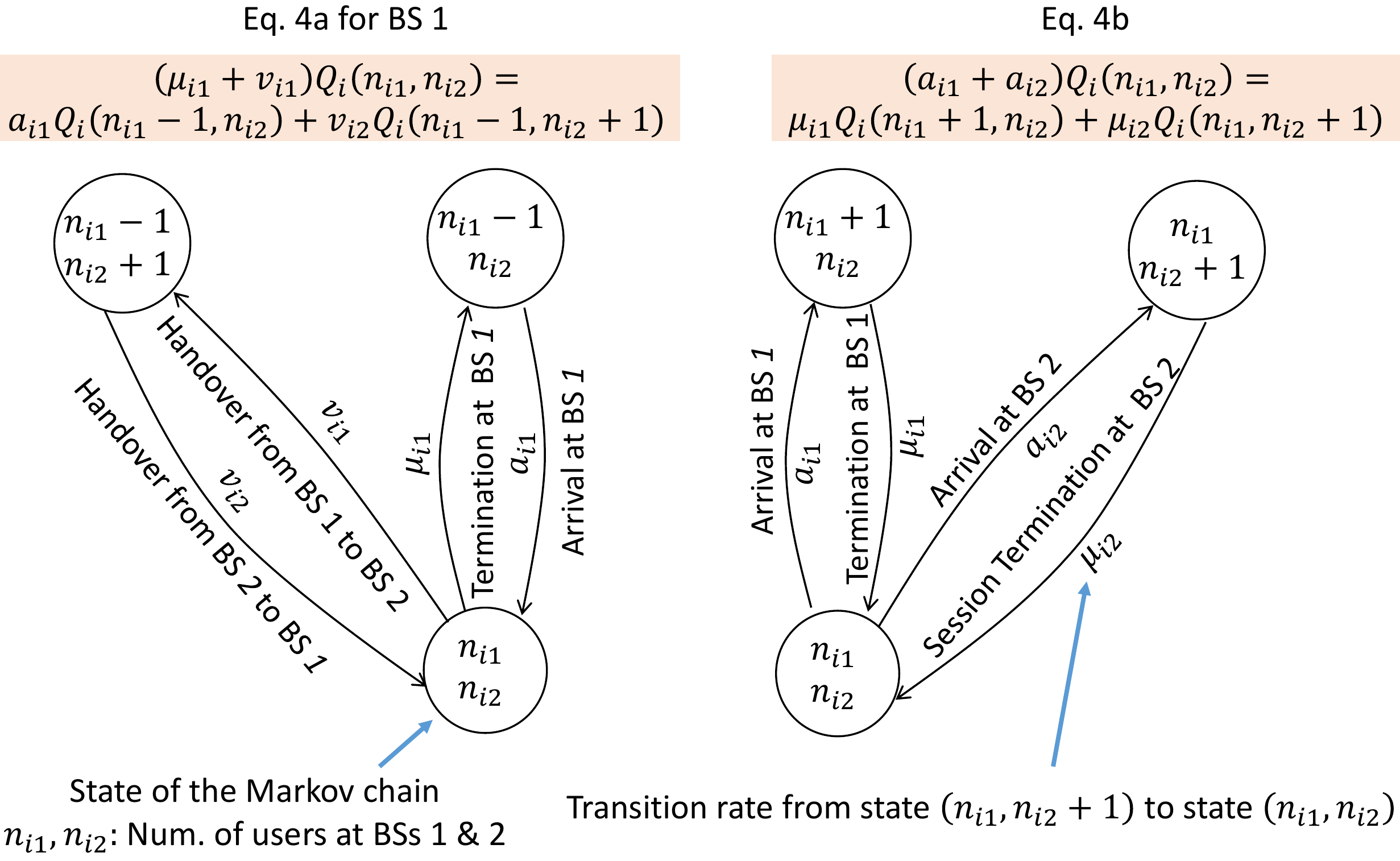}}
\caption{Session arrivals at a BS of the provider $i$ from members of the group $j$ (left). The local-balance equations for a network with two BSs (right).}
\end{figure*}

The mobility of members of the group $j$ in the network of a provider is modeled with a Markov-chain in which a state corresponds to the coverage area of a BS. The total session generation rate of the members of the group $j$ that select the provider $i$ is further divided among its BSs ($k=1,..., K_i$) according to the probabilities $\omega^j_{ik}$. These probabilities correspond to the stationary distribution of the Markov chain modeling the user mobility. Note that the handovers at a BS $k$ of the provider $i$ are modeled with a Poisson process of total rate $v_{ik}$. This rate is estimated according to the fluid flow mobility model \cite{Roy}. We also assume that handovers are performed in a seamless manner.
Table \ref{tbl:technology_parameters} defines the queueing-theoretical parameters for the members of the group $j$ when connected at the network of the provider $i$. Let us now focus on a simple case in which all users select the provider $i$ (i.e., $z_{ji} = 1$ for all $j=1,...,J$). The total session arrival rate at a BS $k$ from members of the group $j$ ($\gamma_{ik}^j$) consists of the new sessions ($a_{ik}^j = \omega_{ik}^j \lambda_j$) and handover sessions from neighbouring BSs (Fig.\ref{fig:traffic_equations}):
%
%
%
\begin{equation} \label{eq:traffic_equations}
\gamma_{ik}^j = a^j_{ik} + \sum_{m=1}^{K_i} \gamma_{im}^jp_{i,m,k}^{(j)}
\end{equation}
The traffic intensity generated by the users of the group $j$ at the BS $k$ of the provider $i$ ($\rho_{ik}^j$) is equal to the ratio of the total session arrival rate at the BS $k$ from members of the group $j$ ($\gamma_{ik}^j$) over the total session departure rate at that BS ($d_{ik}$). However, to characterize the performance of the network of the provider $i$, the total traffic intensity introduced by all user groups at each BS needs to be estimated. By summing the Eq. \ref{eq:traffic_equations} over all user groups, we derive the traffic equations for the network of the provider $i$:
\begin{equation} \label{eq:traffic_equations_total}
\sum^J_{j=1} \gamma_{ik}^j = \sum^J_{j=1} a^j_{ik} + \sum_{m=1}^{K_i} \sum^J_{j=1} \gamma_{im}^jp_{i,m,k}^{(j)}
\end{equation}
We now define the total session arrival rate at the BS $k$ of the provider $i$ from all user groups $\gamma_{ik}=\sum_{j=1}^J \gamma_{ik}^j$, the total arrival rate of new sessions at the BS $k$ of the provider $i$ $a_{ik}= \sum_{j=1}^J a_{ik}^j$, and the average unconditional probability of a handover from the BS $m$ to the BS $k$ over all user groups $p_{i,m,k} = \sum_{j=1}^J \gamma_{im}^jp_{i,m,k}^{(j)} / \gamma_{im}$. The corresponding average conditional probability of a handover from the BS $m$ to the BS $k$ given that a handover occurs is defined as $p^*_{i,m,k} = p_{i,m,k}d_{im}/v_{im}$ Then, the traffic equations can be rewritten as in Eq. \ref{eq:traffic_equations_simple}:
\begin{equation} \label{eq:traffic_equations_simple}
\gamma_{ik} = a_{ik} + \sum_{m=1}^{K_i} \gamma_{im} p_{i,m,k}
\end{equation}
The queueing network of the provider $i$ is modeled as a Markov chain. Each state corresponds to a vector $n_i =(n_{i1}, ...,n_{iK_{i}})$ indicating the number of connected users at all BSs. State transitions correspond to various types of events including session arrivals, terminations, and handovers. The stationary distribution of the Markov chain is estimated by solving the global-balance equations. Such equations set the arrival rate at each state of the Markov chain equal to the departure rate from that state. Due to the Markovian property and the processor-sharing discipline of our system, the global-balance equations can be simplified into a set of local-balance equations \cite{Bolch98}. Unlike global-balance equations, local-balance equations focus on the session arrivals and departures at specific BSs. According to these equations (Eqs. \ref{eq:Local_balance}), the rate leaving a state $n_i$ due to the departure of a user at a specific BS $k$ is equal to the rate entering that state due to the arrival of a user at the BS $k$ either due to a new session or a handover (Eq. \ref{eq:Local_balance_a}). Furthermore, the rate leaving the state $n_i$ due to the arrival of a new session at a BS is equal to the rate entering that state due to the termination of a session at a BS (Eq. \ref{eq:Local_balance_b}).
\begin{subequations} \label{eq:Local_balance}
\begin{multline} \label{eq:Local_balance_a}
d_{ik}Q_i(n_i) = a_{ik}Q_i(n_i - e_{ik}) + \sum_{m=1}^{K_i} v_{im}p_{i,m,k}^*Q_i(n_i - e_{ik}+e_{im})
\end{multline}
\begin{flalign}\label{eq:Local_balance_b}
\text{ }\sum_{k=1}^{K_i} a_{ik}Q_i(n_i) = \sum_{k=1}^{K_i} \mu_{ik} Q_i(n_i + e_{ik}) 
\end{flalign}
\end{subequations}
In Eqs. \ref{eq:Local_balance}, $e_{ik}$ is a vector with all entries equal to $0$ except the $k$-th entry which is equal to $1$. Fig. \ref{fig:Local_balance_equations} illustrates the local-balance equations for a network of two BSs. Given that $\rho_{ik} = \sum_{j=1}^J \rho_{ik}^j < 1$ for each BS of the provider $i$, the stationary distribution of the number of connected users at all BSs can be derived as follows:
\begin{equation} \label{eq:Stationary_chain_distribution}
Q_i(n_i) = \prod_{k=1}^{K_i} \left(1-\rho_{ik}\right)\left(\rho_{ik}\right)^{n_{ik}}
\end{equation}
By substituting Eq. \ref{eq:Stationary_chain_distribution} in the local-balance equations (Eqs. \ref{eq:Local_balance}) and using simple algebra, we derive the traffic equations (Eq. \ref{eq:traffic_equations_simple}). This proves the validity of Eq. \ref{eq:Stationary_chain_distribution}. Given that the stationary distribution is in product form, each BS can be viewed as an independent M/M/1 queue with the processor-sharing discipline. 
%
%

In the general case in which not all users select the provider $i$ (i.e., $z_{ji} < 1$), we can replace $\gamma_{ik}^j$, $a_{ik}^j$, and $\rho_{ik}^j$ with $z_{ji}\gamma_{ik}^j$, $z_{ji}a_{ik}^j$, and $z_{ji}\rho_{ik}^j$, respectively and Eqs. \ref{eq:traffic_equations}-\ref{eq:Stationary_chain_distribution} still hold. In this case, the average number of connected users at the BS $k$ of the provider $i$ is $E[N_{ik}] = \frac{\rho_{ik}}{1-\rho_{ik}}=\frac{\sum_{j=1}^J z_{ji}\rho_{ik}^j}{1-\sum_{j=1}^J z_{ji}\rho_{ik}^j}$ \cite{Kleinrock}, where $\rho_{ik}$ is the traffic intensity introduced by all user groups ($\rho_{ik} = \sum_{j=1}^J z_{ji}\rho_{ik}^j$). When a new user arrives at the BS $k$, it shares the available bandwidth along with all other currently connected users at that BS. Therefore, the amount of bandwidth that a new user gets when it connects to the BS $k$ is $\frac{B_{ik}}{E[N_{ik}]+1} = B_{ik}(1-\sum_{j=1}^J z_{ji}\rho_{ik}^j)$, where $B_{ik}$ is the total bandwidth of that BS. The average data rate as perceived by a user of the group $j$ at the network of the provider $i$ can be computed as the weighted average of the data rate achieved at each BS (Eq. \ref{eq:average_rate}):
\begin{equation} \label{eq:average_rate}
R_{ji}(z) = \sum_{k=1}^{K_i} \omega_{ik}^j B_{ik}(1 - \sum^J_{l=1}z_{li}\rho_{ik}^l)
\end{equation}
%
%
The spatial variability of data rate affects the QoS. Thus, the utility function of the members of the group $j$ (Eq. \ref{eq:Utiltiy_function}) incorporates the average data rate (Eq. \ref{eq:average_rate}) and variance of data rate which is defined as a polynomial of second degree with respect to $z$ (Eq. \ref{eq:variance_data_rate}):
\begin{equation} \label{eq:variance_data_rate}
V_{ji}(z) = \sum_{k=1}^{K_i} \omega^j_{ik}\left(B_{ik}(1-\sum_{l=1}^J z_{li}\rho_{ik}^l) - R_{ji}(z)\right)^2
\end{equation}

The user service selection process employs the average and variance of data rate. The sub-games modeling the user service selection and competition of providers are described in Subsections \ref{subsec:game_of_users} and \ref{subsec:game_of_providers}, respectively. 

\subsection{User service selection} \label{subsec:game_of_users}

The user service selection process is modeled by a population game. Each member of a user group can choose among $I+1$ available strategies $H = \{0,1,...,I \}$. Strategies $1, 2, ..., I$ correspond to subscriptions with the providers $1, 2, ...,I$, respectively, while strategy $0$ denotes the disconnection state. We assume that each group corresponds to a homogeneous sub-population, and as such, the utility attained when selecting a specific strategy is the same for all users in that group. Therefore, it suffices to describe the service selection of the members of the group $j$ with a probability distribution over the set of strategies ($H$). This distribution $z_j = (z_{j0}, z_{j1},...,z_{jI})$ is the {\em strategy profile} of the group $j$ indicating how members of this group are divided among the different strategies (subscriptions and disconnection). The strategy profile of the entire user population consists of the strategy profiles of all groups ($z = (z_{1},...,z_{J})$). Additionally, the {\em market share} that corresponds to a strategy $i$ is the average percentage of customers over all user groups that select this strategy, i.e., $z_i = \sum_{j=1}^J z_{ji}N_j/N$.
 All parameters of a wireless market are defined in Table \ref{tbl:economic_parameters}.
\begin{table}[!t]
\centering
\captionsetup{justification=centering}
\caption{\textbf{Main parameters of a wireless market}}
\begin{tabular}{|c|c|}
\hline \textbf{Parameter} & \textbf{Description}\\
\hline \centering $I$&Number of providers\\
\hline \centering $J$&Number of user groups\\
\hline \centering $N_j$&Number of users in group $j$\\
\hline \centering $N$&Total number of users\\
\hline \centering $c$&Vector with the prices of all providers\\
\hline \centering $H$&User strategies\\
\hline \centering $f_j\left(R_{ji}(z)\right)$&Impact of average data rate on utility function of group $j$\\
\hline \centering $w^j_V$ ($w^j_P$)&Weight of variance of data rate (price) for users of group $j$\\
\hline \centering $u_{ji}\left(z;c\right)$&Utility function of group $j$\\
\hline \centering $d_j$&Average monthly traffic demand in MB of a member of group $j$\\
\hline \centering $z(t)$&User strategy profile at time $t$\\
\hline \centering $z^*(c)$&Equilibrium of users when price vector is c\\
\hline \centering $P$&Providers\\
\hline \centering $C$&Provider strategy profiles\\
\hline \centering $\sigma_i(c)$&Utility function of provider $i$\\
\hline
\end{tabular}
\label{tbl:economic_parameters}
\end{table}

{\em Utility function of the group $j$.} A user from the group $j$ selects a strategy (i.e., a subscription or disconnection) based on the average and spatial variance of the achievable data rate at the networks of providers and the offered prices:
\begin{equation} \label{eq:Utiltiy_function}
u_{ji}(z;c) =
\begin{cases}
f_j\left(R_{ji}(z)\right) - w^j_V V_{ji}(z) - w^j_Pc_i(d_j)&\text{if }i = 1,...,I \\
0 & \text{if }i = 0
\end{cases}
\end{equation}
The function $f_j$ is concave, strictly increasing, and non-negative and defines the impact of the average data rate ($R_{ji}(z)$) on the utility of the members of the group $j$. Such functions have been frequently used in the literature (e.g., \cite{Gagic14, Kelly98}). In the analysis, we assume that $f_j$ is exponential, i.e., $f_j(x) = w^j_R(\tau_j-\exp(-h_jx))$. Its main characteristic is that it defines a diminishing return for users when the QoS improves\footnote{Other utility functions, e.g., logarithmic and isoelastic ones, could be also employed.}. The parameter $w^j_R$ expresses the willingness to pay of the members of the group $j$. The larger the $w^j_R$, the larger the maximum price that users from the group $j$ can pay. The parameter $h_j$ defines the sensitivity of the members of the group $j$ to low data rate. The larger the $h_j$, the larger the tolerance of users to a low data rate. The impact of the variance of data rate ($V_{ji}(z)$) and price of the subscription of the provider $i$ ($c_i(d_j)$) is assumed to be linear and their significance is indicated by the positive weights $w^j_V$ and $w^j_P$, respectively. The price that the members of the group $j$ pay when selecting the subscription with the provider $i$ depends on the average traffic demand they produce in a period of a month ($d_j$) according to Eq. \ref{eq:pricing}:
\begin{equation} \label{eq:pricing}
c_i(d_j) =
\begin{cases}
c_{i1}&\text{if }0 < d_j \leq D_1 \\
c_{i2}&\text{if }D_1 < d_j \leq D_2 \\
...\\
c_{iS_{i}} & \text{if } D_{S_{i}-1} < d_j \leq D_{S_{i}}
\end{cases}
\end{equation}

{\em Dataplans.} Most of the operators employ pricing strategies based on the data instead of talk time or text, in response to the changes in the usage patterns (e.g., \cite{ramn16}). The provider $i$ offers $S_i$ distinct dataplans each for different traffic demand levels. Depending on which interval the traffic demand of a user lies, that user pays a different price. In other words, the provider $i$ charges each user with a flat rate that depends on its level of traffic demand according to Eq. \ref{eq:pricing}. We assume that each user is aware of its own average traffic demand per month when selecting a service \footnote{In this paper, the term ''service" refers to a subscription with a provider or the disconnection, while the term ''dataplan" refers to the offered price of a provider for a specific interval of traffic demand. A user selects a service and pays the price corresponding to its level of traffic demand.}. Furthermore, when a user selects the disconnection (i.e., $i = 0$), it attains utility equal to $0$. The parameters of the utility function $u_{ji}$ (i.e., $w^j_R$, $\tau_j$, $h_j$, $w^j_V$, $w^j_P$ and $d_j$) constitute the profile of the members of the group $j$.

{\em Evolution of user-decision making.} Based on the utility functions of all user groups, the evolution of the strategy profile of users ($z(t)$) is described by the Logit dynamics, a system of ordinary differential equations (Eq. \ref{eq:Logit_dynamics}). Compared to other population dynamics, the Logit dynamics have two attractive properties: (1) They can capture the user bounded rationality and stickiness/loyalty to certain providers. (2) They are innovative dynamics.
\begin{equation} \label{eq:Logit_dynamics}
\frac{dz_{ji}(t)}{dt} = \frac{r}{1+\sum_{k\neq i} \exp \left( \frac{u_{jk}(z(t)) - u_{ji}(z(t))}{\epsilon}\right)}-rz_{ji}(t)
\end{equation}
%
Note that we assume that the evolution of the user decisions manifests at a much slower time scale compared
to the time scale at which sessions arrive and depart at the BSs of a provider. Specifically,
the session arrivals and departures are performed at a time scale of minutes, while the user
decisions manifest at a time scale of days or even months. Therefore, when the user strategy
profile changes from $z$ to $z_0$ over the course of several days, there is enough time for
the queueing network of a provider to reach the equilibrium.
We can then
assume that the average data rate changes from $R_{ji}(z)$ to $R_{ji}(z_0)$ and the spatial variance
of data rate changes from $V_{ji}(z)$ to $V_{ji}(z_0)$.
The parameter $r$ controls the speed of the dynamics, while $\epsilon$ is the noise and takes values in the interval $[0, \infty)$. When $\epsilon = 0$, users are completely ``rational'', i.e., always select the strategy that maximizes their utility function. Specifically, if the strategy $k$ has slightly larger utility than strategy $i$ that has been selected by a rational user, this rational user will then switch to strategy $k$. However, this behaviour is not realistic. 

Users are reluctant to switch to another provider when the additional benefits of switching are small. According to a survey study, users should be offered additional benefits of around $40$\% before they are highly likely to change their provider \cite{Xavi08}. Several aspects, such as, the brand name and brand equity \cite{Sven13}, reputation of a provider \cite{Kim04, Eshg07}, market share of a provider \cite{Kara13}, length of customer association with a provider \cite{Seo08}, and force of habit \cite{Lin06}, affect the user-decision making. In this work, the noise of the Logit dynamics models the aggregate effect of those aspects. As the noise increases from $0$ towards infinity, users become ``stickier'' with their selected service and change provider only when the benefits in terms of price and QoS are large enough. 

To compute the equilibrium of users for a given set of prices offered by the providers, we solve the system of the Logit dynamics (Eq. \ref{eq:Logit_dynamics}) using a standard ODE solver starting from an initial point at which all user groups are uniformly distributed across the available strategies (i.e., subscriptions with providers and disconnection). The point at which the Logit dynamics converge is the user equilibrium.

\subsection{Competition of providers} \label{subsec:game_of_providers}

The competition of providers is modelled as a normal-form game $\left(P, C, \{\sigma_i\}_{i \in P}\right)$. In this game, each provider (say provider $i$) offers $S_i$ distinct dataplans to users. The strategy of that provider $c_i = (c_{i1}, ...,c_{iS_{i}})$ is a vector containing the prices of all those dataplans. Each price is selected from a closed interval $[0,\text{ } C^{max}_{i}]$. The strategy space of providers is the set of all possible combinations of prices that can be offered in the market and is a rectangle of the form $C = [0,\text{ }C^{max}_1]^{S_{1}}\times[0,\text{ }C^{max}_2]^{S_{2}}\times...\times[0,\text{ }C^{max}_I]^{S_{I}}$. Each point of the strategy space $c = (c_1,...,c_I)$ is a vector containing the offered prices of all providers. 

We assume that each provider can model the user population at different levels of detail. They analyze customer data, e.g., demographic information as well as information about their traffic demand and profile, to perform market segmentation. In the performance evaluation, we considered a synthetic user profiles dataset (for the entire customer population) and that each provider can extract information for a different number of market segments (user clusters), corresponding to an incomplete view of the market.
\footnote{Obviously the analysis can be extended considering that providers have access to different datasets of the customer population.}
The larger the number of market segments, the more detailed information about the user population.

Each provider, after estimating the market segments of users, it models their decision making based on the system of Logit dynamics (Eq. \ref{eq:Logit_dynamics}). For each set of offered prices $c$, a provider solves the system of Logit dynamics modeling the decision making of these clusters of users starting from uniform initial conditions and computes the corresponding equilibrium point ($\hat{z}^*(c)$). Based on this equilibrium, the utility function of the provider $i$ is defined according to Eq. \ref{eq:provider_utility_fun}:
\begin{equation} \label{eq:provider_utility_fun}
\sigma_i(c) = \sum_{s=1}^{S_{i}} \sum_{j: D_{s-1} < d_j \leq D_{s}} N_j \hat{z}^*_{ji}(c)c_{is}
\end{equation}
This function computes the sum of the revenue collected by each of the offered dataplans of the provider $i$. For a dataplan $s$, only the payments of the clusters corresponding to this dataplan are considered (i.e., the user clusters that produce an average traffic demand lying in the interval $(D_{s-1}, D_s]$).
%

\begin{figure*}[t!]
\centering
\subfloat[]{\label{noise_0_provider1}\includegraphics[width=0.35\textwidth, height=35mm]{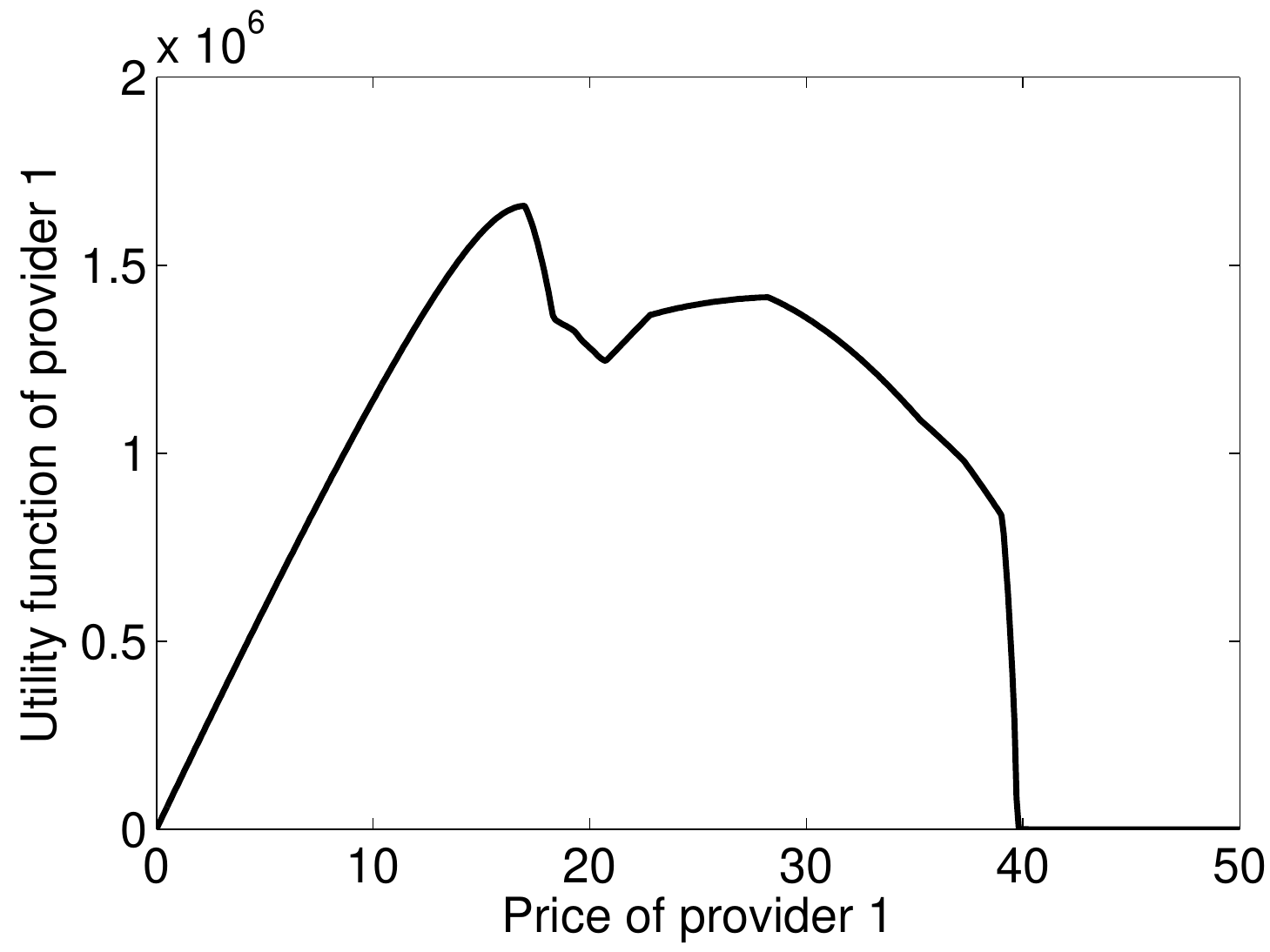}}
\subfloat[]{\label{noise_5_provider1}\includegraphics[width=0.35\textwidth, height=35mm]{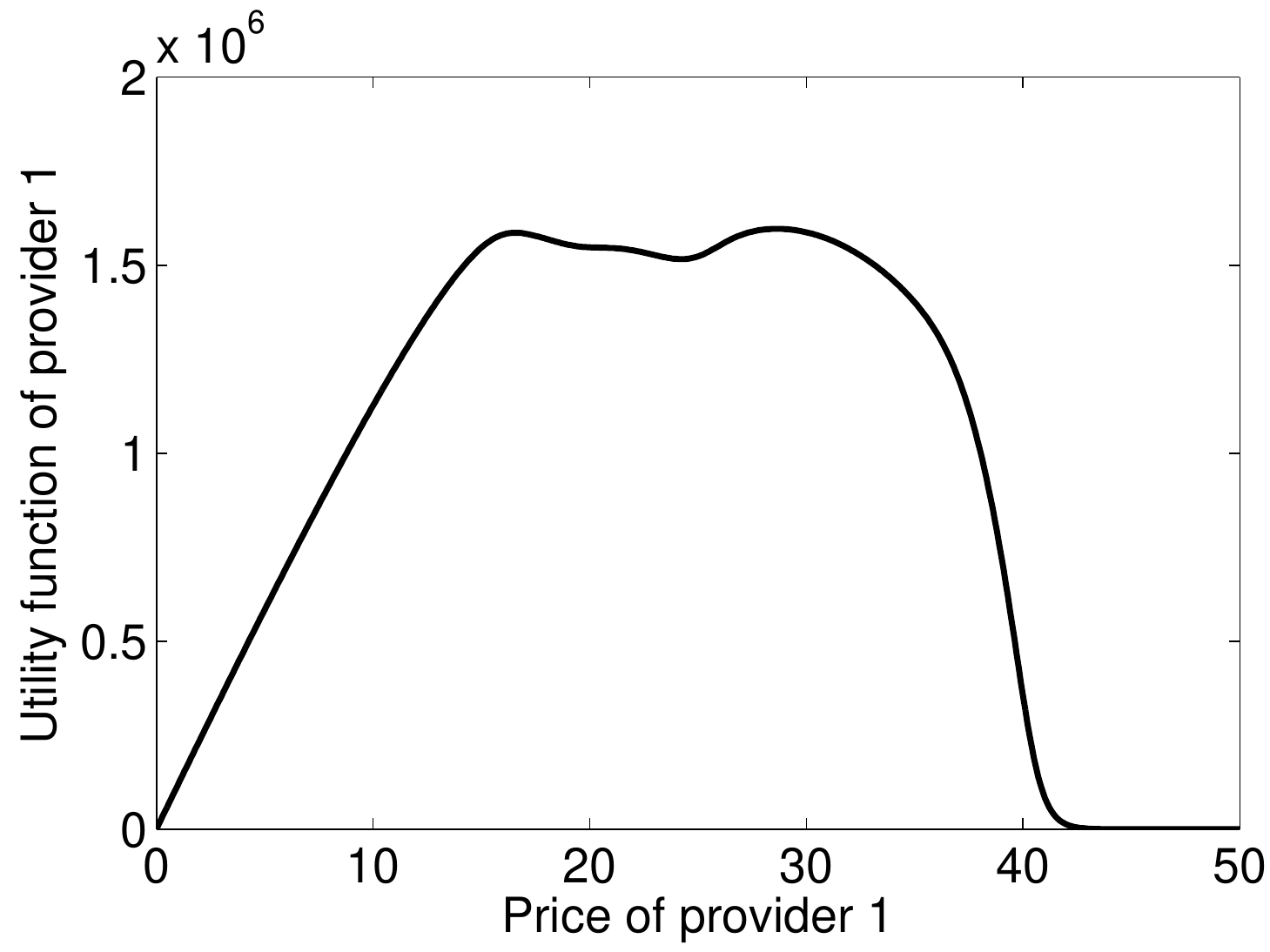}}
\subfloat[]{\label{noise_15_provider1}\includegraphics[width=0.35\textwidth, height=35mm]{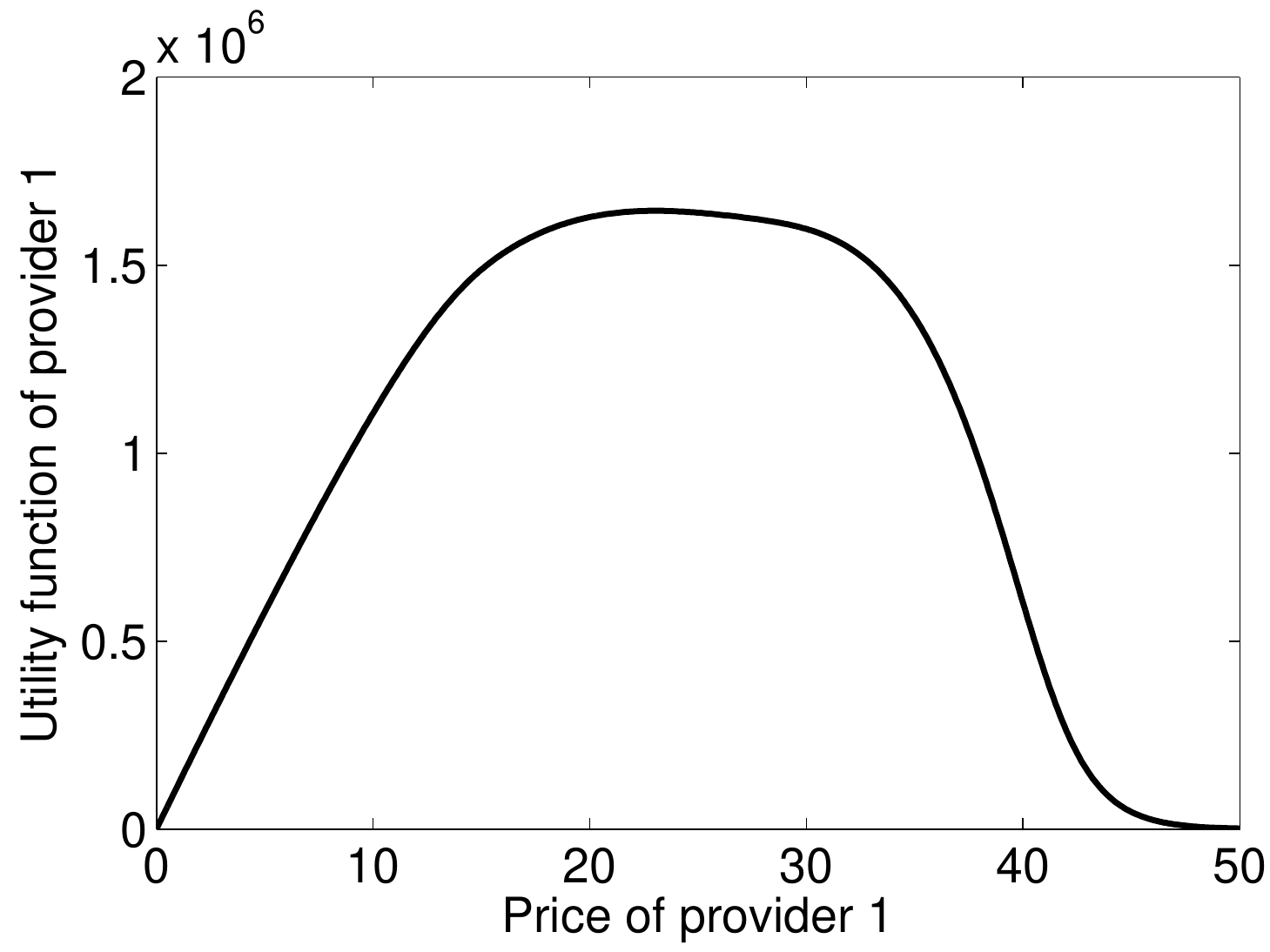}}
\caption{Effect of noise $\epsilon$ on the utility function of a provider in a market with 5 user groups, $\epsilon = 0$ (left), $\epsilon = 0.5$ (middle), and $\epsilon = 1.5$ (right).}
\label{fig:noise}
\end{figure*}

When a user is rational (i.e., $\epsilon = 0$), it selects the provider that maximizes its utility function. Let us assume that such a user has selected the provider $i$. If at a certain point the subscription of another provider $k$ becomes slightly more profitable compared to the subscription of the provider $i$, the user will switch providers. This extreme behaviour of rational users introduces technical difficulties in the analysis: Various discontinuities appear in the derivatives of the utility functions of providers. We developed a methodology to analytically compute the NEs of providers under such discontinuities. This methodology divides the strategy space of providers into various subsets called ``regions'' in which the provider utility functions are continuously differentiable and analyses the game of providers at each one of those regions separately. At the final step, it combines the results of the analysis from all the regions to compute the global NEs of providers. A more detailed description of this methodology can be found in our earlier work \cite{forte0215}. 

This modeling approach can effectively compute the NEs of providers, when users are completely rational and providers model users at the macroscopic level (i.e., considering only $1$ cluster) and offer only $1$ dataplan\footnote{In the worst case, the algorithm needs to solve two non-linear systems of $I$ equations and a system of $2I$ non-linear inequalities.}. However, what happens when providers have a more rigorous knowledge about the market segments? In such a case, the computation of the global NE becomes challenging: As the number of user clusters increases, the number of (hyper-)surfaces at which the utility functions of providers have discontinuous derivatives in the strategy space of providers also increases.

This issue is mainly due to the assumption of the user rationality. Rationality is  a set of principles that describes the utility maximinzing choice. Under this assumption, users can drastically change their behaviour even under small deviations in their utility function. To more realistically model the user behaviour, we use the Logit dynamics (Eq. \ref{eq:Logit_dynamics}). Among other standard population dynamics, such as, the replicator, best response, BNN, and Smith dynamics, these are the only ones that can model the user irrationality \cite{Sand15}. As explained earlier, their noise parameter ($\epsilon$) indicates how much the user decisions deviate from the optimal ones based on their utility function capturing the effect of other factors that influence users (e.g., psychological/social factors).  In general, issues such as the price-quality correlation, impact of objective information and personal experience on user strategy have been examined in marketing e.g., \cite{tell90}. 

The noise of the Logit dynamics does not only make the model of users more realistic but also simplifies the analysis by ``smoothing out'' the discontinuities in the derivatives of the utility functions of providers. An example of the effect of noise is shown in Fig. \ref{fig:noise}. This figure presents the utility function of a provider (say provider $1$) offering only $1$ dataplan in an oligopoly with $5$ distinct user groups, when the prices of its competitors remain fixed. When the noise is equal to zero (i.e., $\epsilon = 0$), the rational behaviour of users results in various discontinuities in the derivative of the utility function of the provider $1$ (Fig. \ref{noise_0_provider1}). When the value of noise slightly increases (i.e., $\epsilon = 0.5$), the discontinuities are smoothed out (Fig. \ref{noise_5_provider1}). If the noise increases a little further (i.e., $\epsilon = 1.5$), the utility function of the provider becomes concave. This function (Fig. \ref{noise_15_provider1}) does not deviate significantly from the corresponding one when users are completely rational (Fig. \ref{noise_0_provider1}).

In each market case, by selecting a sufficient amount of noise, the utility functions of providers become concave. This simplifies the estimation of a global NE significantly. One should simply set the derivatives of the utility functions of providers with respect to their prices equal to $0$:
\begin{equation} \label{eq:provider_equilibrium}
\frac{\partial \sigma_i(c)}{\partial c_{is}} = 0, \text{ for all } i = 1, \ldots, I \text{ and } s = 1, \ldots, S_{i}
\end{equation}
\noindent To compute the Nash equilibrium of providers, we solve the system of non-linear Eqs. \ref{eq:provider_equilibrium} using a standard numerical-analysis algorithm. When the utility functions of providers are concave, a solution of the system \ref{eq:provider_equilibrium} ($c^*$) is guaranteed to be a global NE of the game of providers. At this point, the utility functions of providers are maximized given the prices of their competitors \cite{Boyd04} and therefore, no provider has the incentive to change its strategy. Our algorithm for the estimation of a NE proceeds as follows: First the system of Eqs. \ref{eq:provider_equilibrium} is solved. If a solution is reported, it will be then verified whether or not it corresponds to a global NE (i.e., at which the utility function of each provider is maximized given the prices of the other providers). This final verification step is necessary given that the concavity of the utility functions of providers is not always guaranteed. For example, if the selected value of noise is not sufficiently large, the utility function of one or more providers may not be concave. In such a case, for a point to be   a Nash equilibrium, the system of Eqs. \ref{eq:provider_equilibrium} is still necessary but not sufficient. Therefore, a solution of Eqs. \ref{eq:provider_equilibrium} could correspond to a local maximum of the utility functions of providers, i.e., a ``local NE'' instead of a global one.

\section{Performance evaluation} \label{sec:Perf_evaluation}

We implemented the modeling framework in Matlab and instantiated a wireless access market of a small city, represented by a rectangle of 14.4 km x 12.5 km. 
This market includes 4 providers and a population of $300,000$ users.
\begin{figure*}[t!] 
\centering
\subfloat[]{\label{fig:scenario_1}\includegraphics[width=0.35\textwidth, height=40mm]{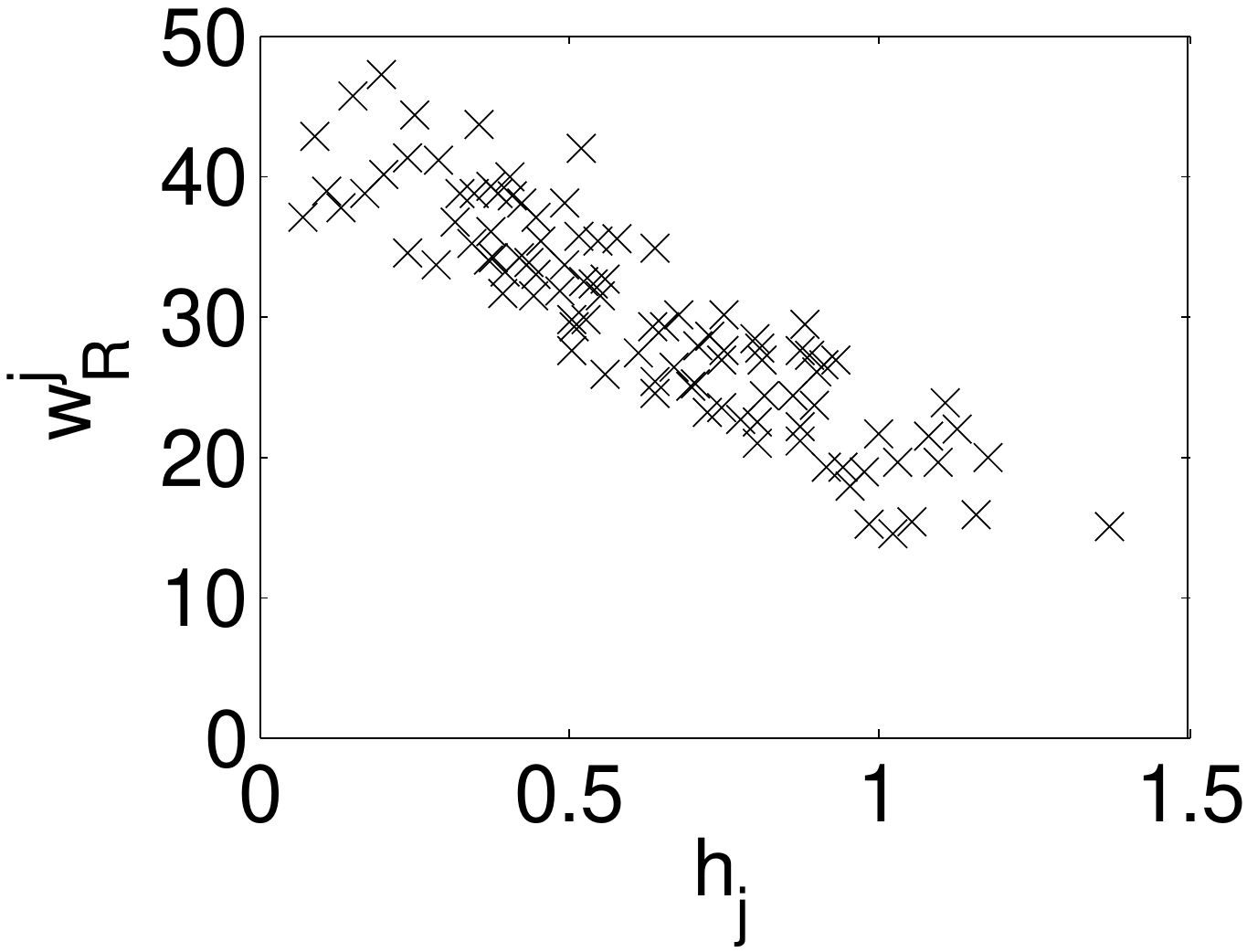}}
%
\subfloat[]{\label{fig:scenario_2}\includegraphics[width=0.35\textwidth, height=40mm]{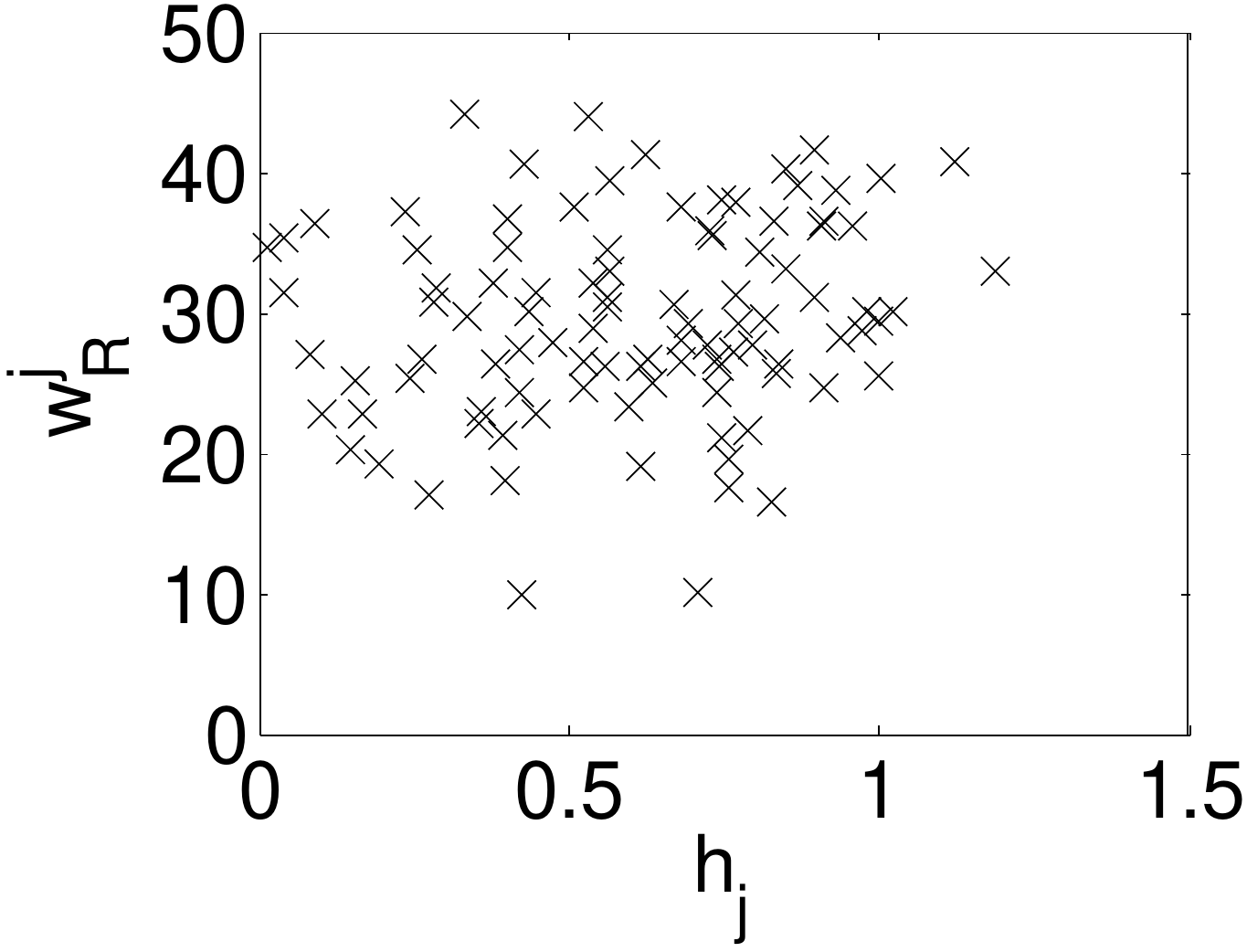}}
\subfloat[]{\label{fig:User_population}\includegraphics[width=0.35\textwidth,height=40mm]{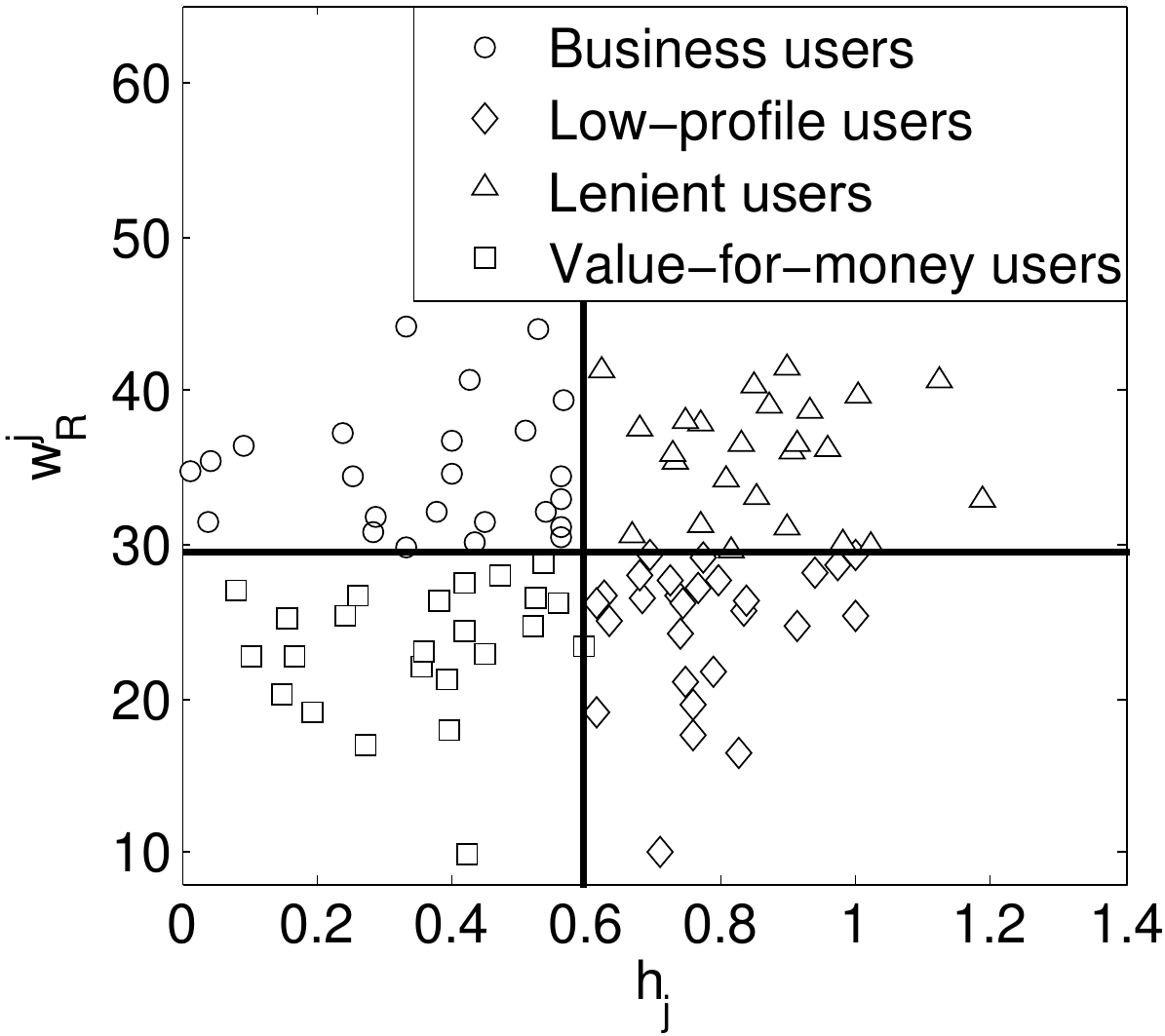}}
  \caption{User group distribution when $w_R^j$ and $h_j$ are correlated (a) and when they are independent (b), respectively. Distribution of user groups for independent $w^j_R$ \& $h_j$ (c).}
\label{fig:scenarios}
\end{figure*}
 Each provider has deployed a cellular network covering the entire city. The BSs at each network are placed on the sites of a triangular grid, with a distance between two neighbouring sites of 1.6 km. 
The maximum data rate with which a BS can serve sessions is $25$, $22$, $19$, and $16$ Mbps for the providers 1, 2, 3, and 4, respectively.
The provider 1 is the strongest provider in the market with the largest capacity at its BSs (i.e., 25 Mbps), while the provider 4 is the weakest one with the lowest capacity (i.e., 16 Mbps). 
The average size of a session is $10$ MB. Furthermore, the session service rate of a BS is $\mu_1 = 18.75$, $\mu_2 = 16.50$, $\mu_3 = 14.25$, and $\mu_4 = 12.00$ sessions/min for the providers 1, 2, 3, and 4, respectively. We also consider different cases for the user traffic demand (i.e., an average user session generation rate from $0$ up to $1.5$ sessions/hour) and analyse its impact on providers and users.

\subsection{Modeling users at different levels of detail} \label{sec:heterogeneity}

\begin{figure*}[t!] 
  \centering
\subfloat[]{\label{corr_disc}\includegraphics[width=0.35\textwidth, height=42mm]{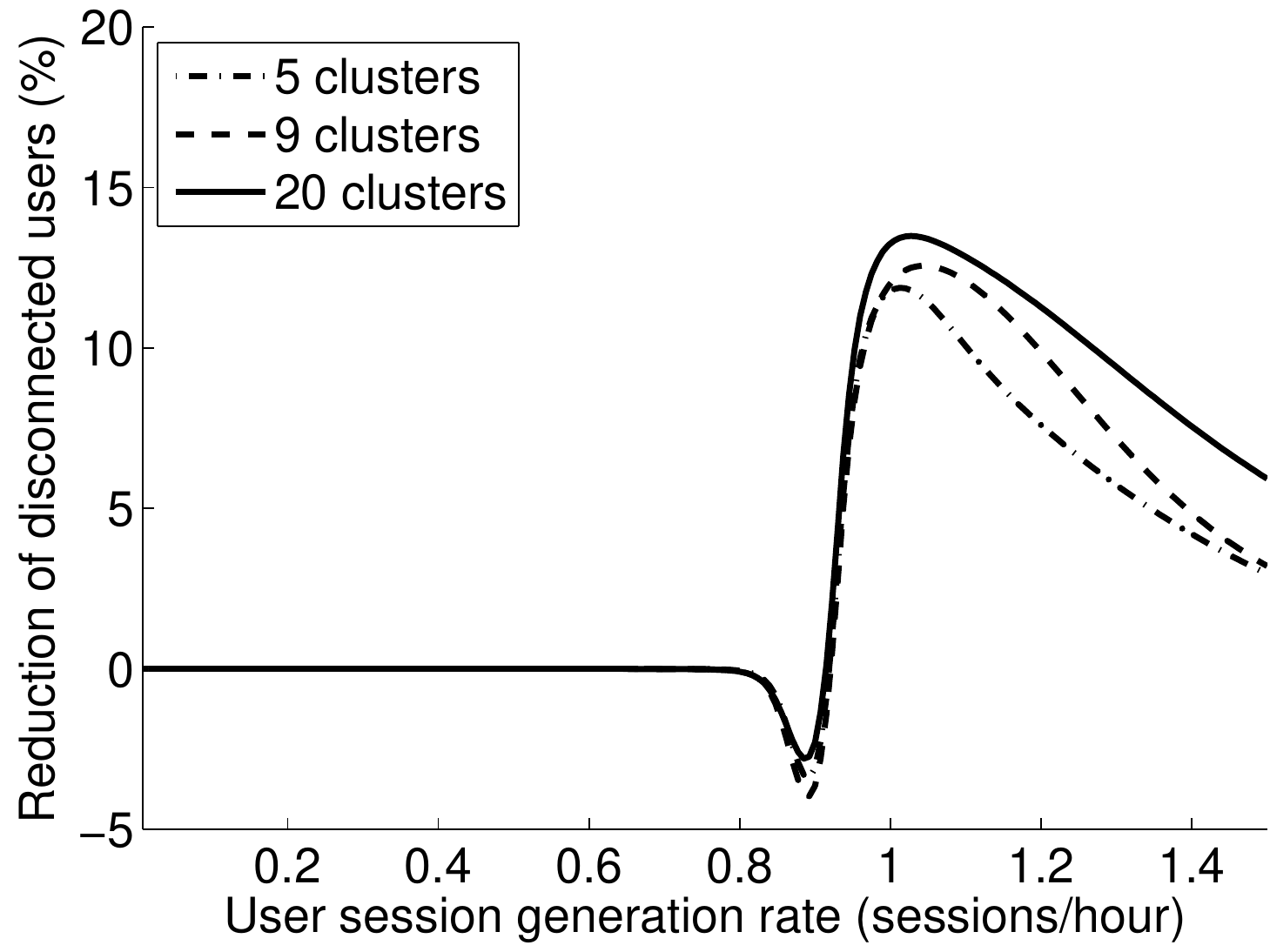}}
\subfloat[]{\label{corr_provider_1}\includegraphics[width=0.35\textwidth, height=42mm]{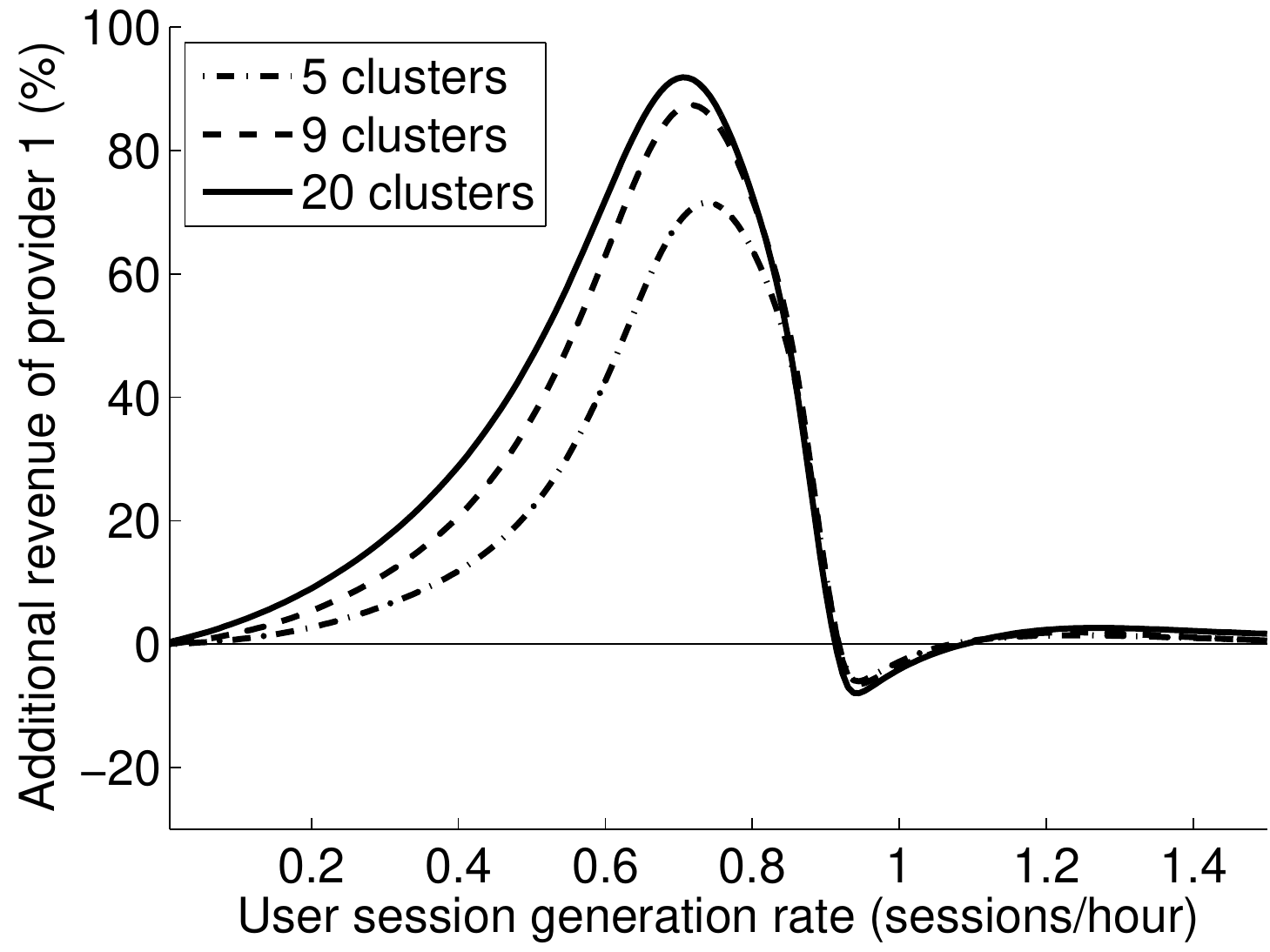}}
\subfloat[]{\label{corr_provider_4}\includegraphics[width=0.35\textwidth, height=42mm]{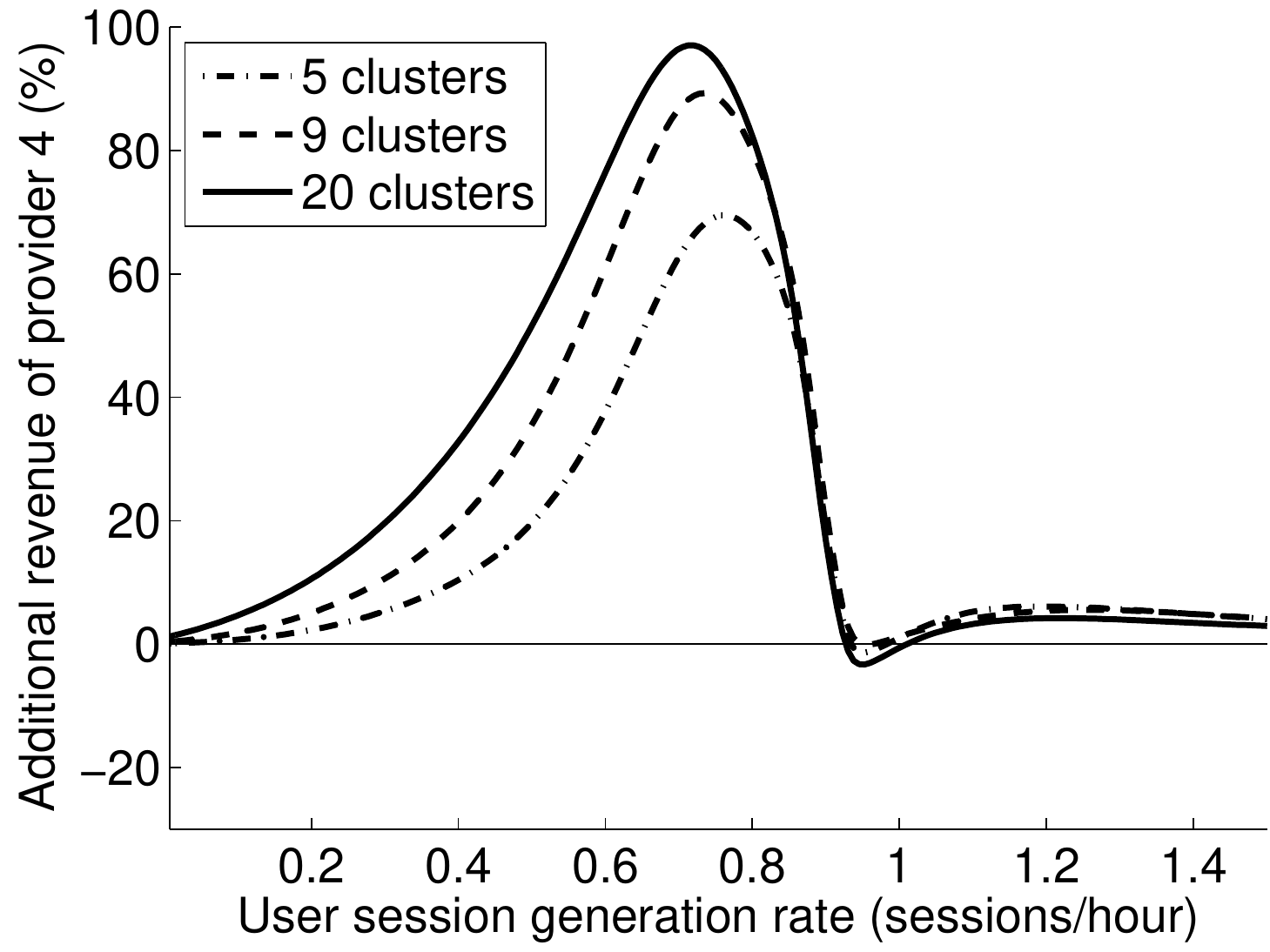}}\\
\subfloat[]{\label{ind_disc}\includegraphics[width=0.35\textwidth, height=42mm]{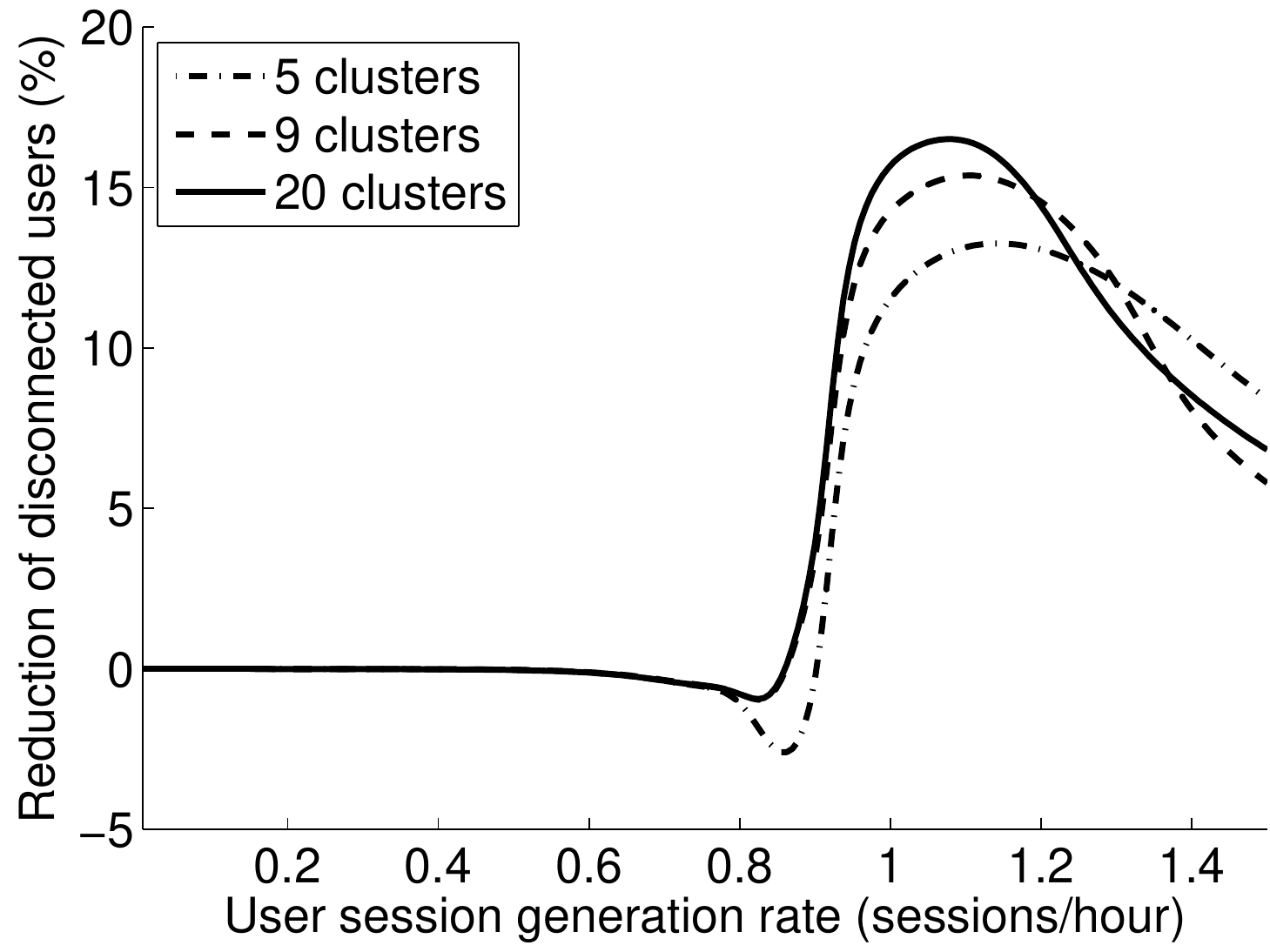}}
\subfloat[]{\label{ind_provider_1}\includegraphics[width=0.35\textwidth, height=42mm]{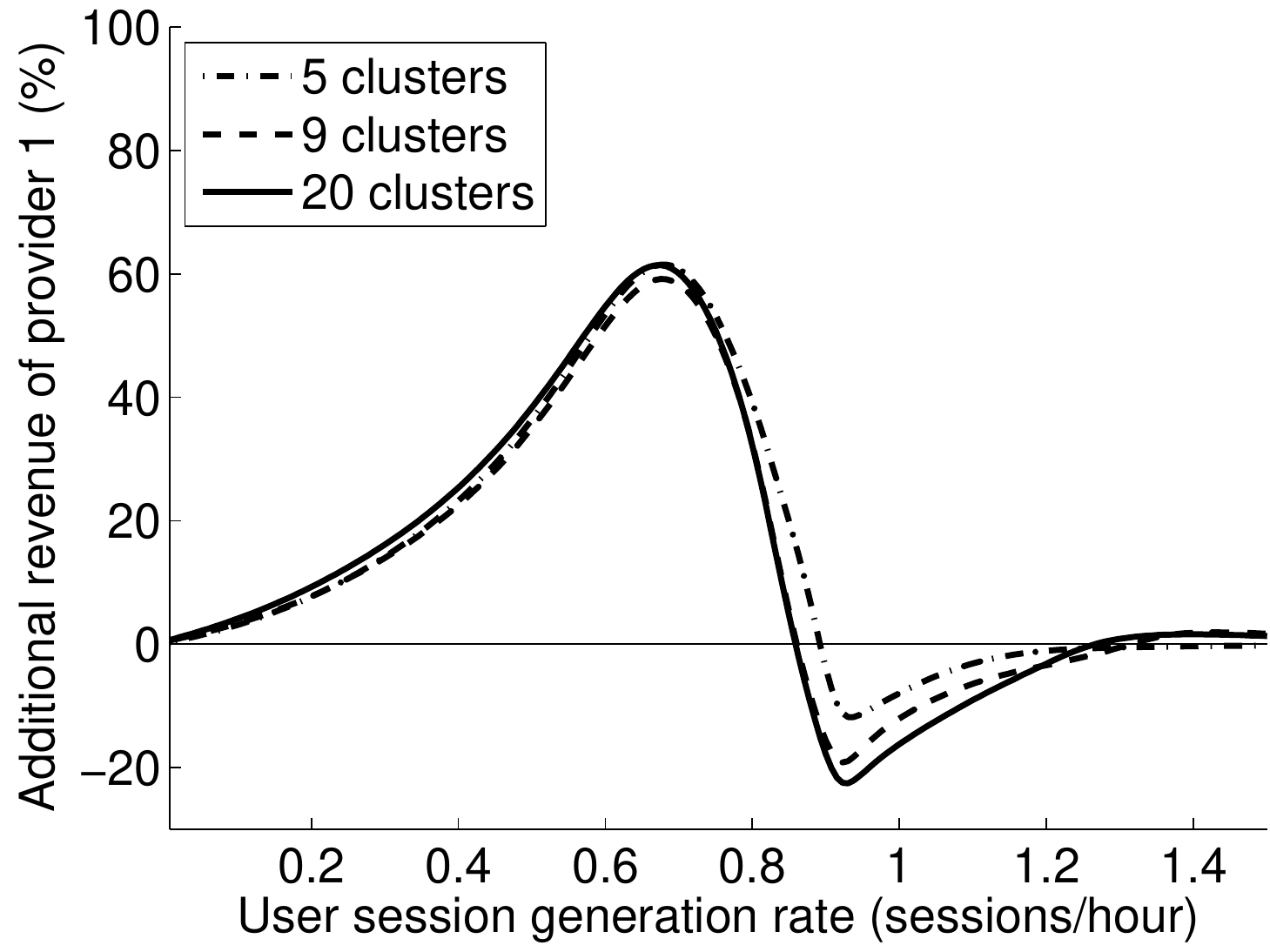}}
\subfloat[]{\label{ind_provider_4}\includegraphics[width=0.35\textwidth, height=42mm]{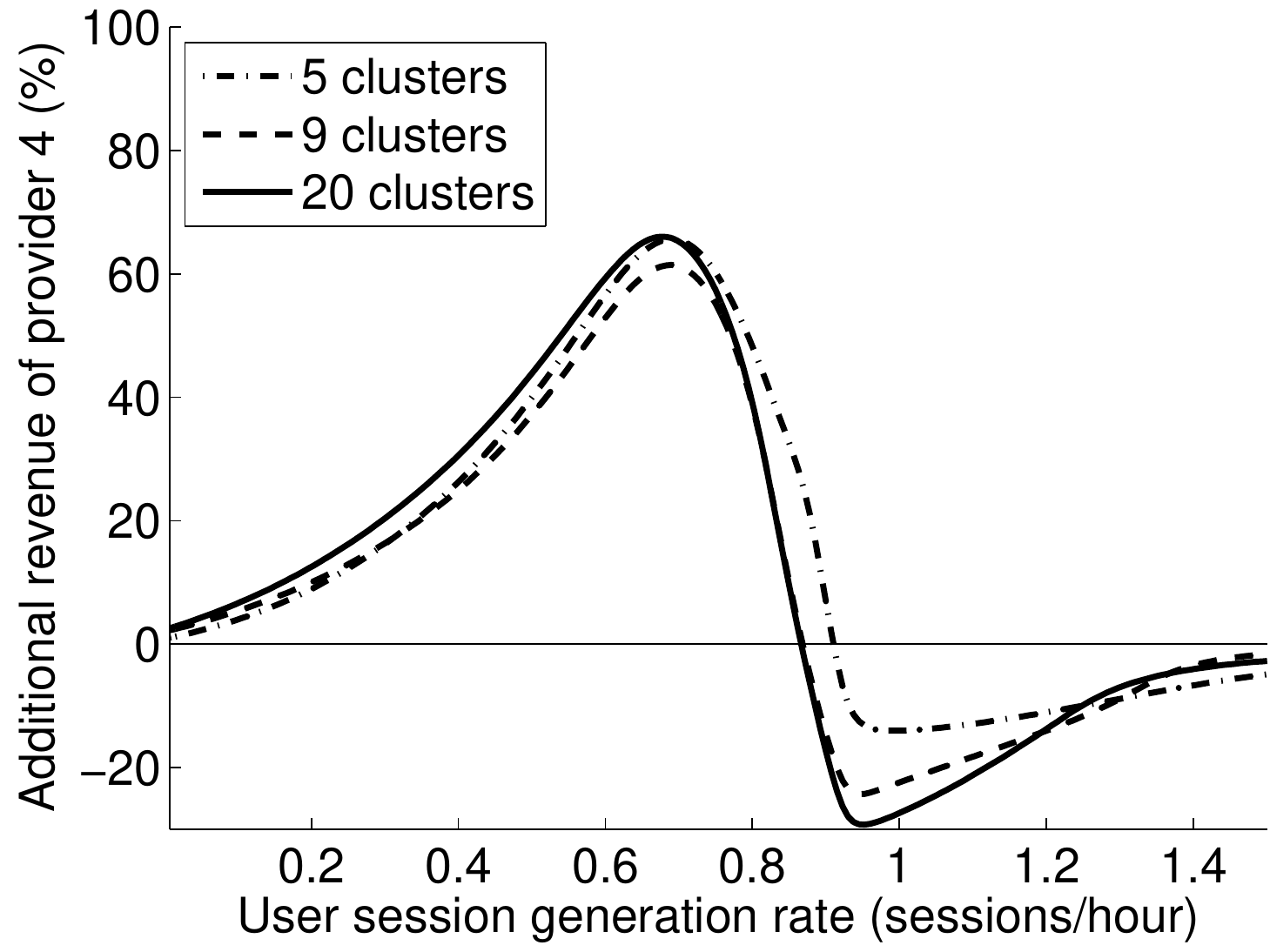}}
  \caption{Performance gains when providers model users at different levels of detail compared to macroscopic modeling when $w_R^j$ and $h_j$ are correlated (top) and when they are independent (bottom), respectively.}
  \label{fig:multi_layer}
\end{figure*}

We consider a heterogeneous user population consisting of $100$ distinct groups, which represents the most detailed picture of the user population. To model a diverse user population, we selected
the maximum willingness to pay ($w^j_R$) and tolerance on low data rate ($h_j$) of these groups follow normal distributions of mean $30$ and $0.6$ and 
standard deviation of $7.6$ and $0.3$, respectively. 
In real markets, those two parameters are often correlated. 
For example, users with a large willingness to pay (high $w_R^j$) usually demand a high QoS and are less tolerant on low data rate (low $h_j$). 
This is due to the fact that the service charge creates an expectation for the perceived quality. 
However to highlight the impact of the correlation between the user willingness to pay and tolerance on low data rate, we also defined a scenario in which those two parameters are independent. 
Specifically, in the first (main) scenario, the cross correlation of $w_R^j$ and $h_j$ is equal to $-0.85$ (Fig. \ref{fig:scenario_1}). 
This means that users with a large willingness to pay (high $w_R^j$) usually are less tolerant on low data rate (low $h_j$). In the second scenario, 
the cross correlation of $w_R^j$ and $h_j$ is equal to $0$ and the maximum willingness to pay and data rate requirements of groups are completely independent (Fig. \ref{fig:scenario_2}). 
In both scenarios, the traffic demand ($d_j$) is the same for all user groups and providers offer only one dataplan. 
Additionally, the weight of data rate variability ($w^j_V$) is set equal to $0$ for all user groups, while the noise value ($\epsilon$) is set equal to $1.5$.
The noise parameter should reflect the user behaviour. However given the lack of real-world datasets,
the minimum value of noise that results in concave utility functions (for the providers) was selected. 

We distinguish four distinct user categories the {\em business users}, {\em low profile users}, {\em value for-money users}, and {\em lenient users}. Business users have a high willingness to pay (i.e., high value of $w^j_R$) but are highly sensitive on low data rate (i.e., low value of $h_j$). Low-profile (or basic) users are the opposite: They can not afford a high price but are more tolerant on low data rates. Value-for-money (or best-deal) users
are the most demanding ones in the market. They have a low
willingness to pay and can not tolerate low values of data rate.
Finally, lenient users are characterized by a high willingness
to pay and high tolerance in low data rate. These users
make their decisions considering mainly the price, i.e., they
search for the cheapest service. Such consumer classification in terms of choice strategies is common in marketing, especially when price is better known than quaility \cite{tell90}.
We clustered the 100 sub-populations to the above four categories. 
The characteristics of these
four categories are given in Table \ref{table:user_groups_cats}.

\begin{table}[!t]
\centering
\captionsetup{justification=centering}
\caption{\textbf{The four main user categories}}
\begin{tabular}{|c|c|c|}
\hline \textbf{Category} & \textbf{Willingness to} & \textbf{Tolerance on low}\\
& \textbf{pay ($w^j_R$)}&\textbf{data rate ($h_j$)}\\
\hline \centering Business users&$>$ 50\% percentile&$<$ 50\% percentile\\
\hline \centering Low-profile users&$<$ 50\% percentile&$>$ 50\% percentile\\
\hline \centering Value-for-money users&$<$ 50\% percentile&$<$ 50\% percentile\\
\hline \centering Lenient users&$>$ 50\% percentile&$>$ 50\% percentile\\
\hline
\end{tabular}
\label{table:user_groups_cats}
\end{table}

For each scenario, we have defined different market cases in which providers model the user population at different degrees of detail when estimating their utility functions. Each provider applies a clustering algorithm (e.g., K-means) on the profiles of the $100$ user groups and determines a set of $5$, $9$, or $20$ {\em representative user clusters}. Providers consider these sets of clusters when estimating their utility functions (according to Eq. \ref{eq:provider_utility_fun}). As the number of clusters increases, the modeling of the user population becomes more accurate but the computational complexity of estimating the NE of providers increases. Our objective is to select the most appropriate level of detail that results in large performance gains for providers at a low computational complexity. Fig. \ref{fig:multi_layer} shows the {\em additional} benefits of users and providers obtained when providers model the user population at different levels of detail compared to modeling users macroscopically (i.e., with only $1$ cluster) when $w_R^j$ and $h_j$ are correlated (Figs. \ref{corr_disc} - \ref{corr_provider_4}) and when they are independent (Figs. \ref{ind_disc} - \ref{ind_provider_4}), respectively. 

{\em Case a:} When the maximum willingness to pay ($w^j_R$) and tolerance on low data rate ($h_j$) are correlated, modeling the user population at a higher level of detail (with a larger number of clusters) pays off for providers and users. Providers increase their revenue in almost all cases (Figs. \ref{corr_provider_1} and \ref{corr_provider_4}) and the percentage of disconnected users is significantly reduced (Fig. \ref{corr_disc}). When providers model users at the macroscopic level, they apply a {\em homogeneous marketing} strategy by selecting their prices considering only aggregate profiling information for the entire population, without detailed information for the different sub-populations that reflect the diversity of the market. The provider $1$ offers a price that is close to the prices of the other providers and attracts both business and low-profile users. Additionally, the providers $3$ and $4$ offer relatively high prices losing customers and revenue form the low-profile user groups. In such a market, except from the revenue of providers, the performance of users is also sub-optimal. A large percentage of users ends up selecting the provider $1$ resulting in a degradation of the offered quality of service. Furthermore, due to the relatively high prices of the providers $3$ and $4$, more low-profile users become disconnected. In other words, under a macroscopic view of the market, providers make suboptimal decisions, which have a negative impact on their revenue and performance of users. 

\begin{figure} [t]
\centering
\includegraphics[width=4in]{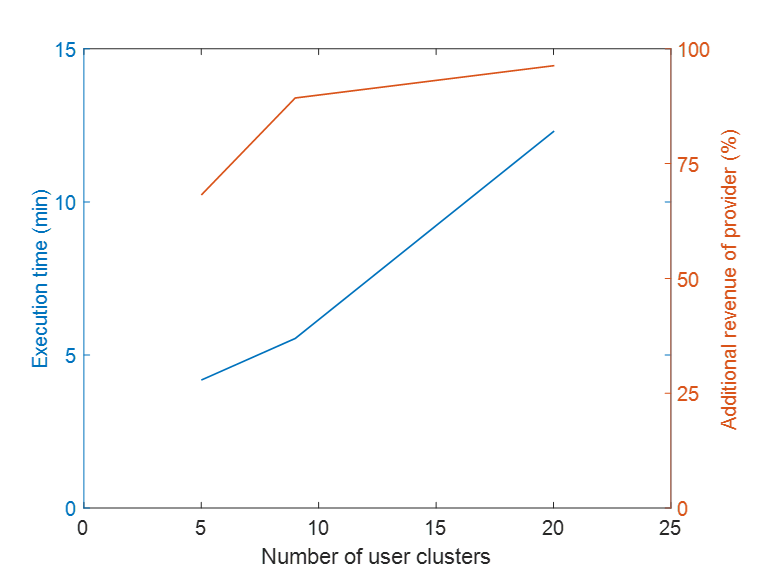}
\caption{Execution time needed to perform the market analysis and additional revenue gain of a provider at different levels of detail.} 
\label{fig:execution}
\end{figure}

By employing a larger number of clusters for the estimation of their utility functions (e.g., $5$, $9$ or $20$ clusters), providers obtain a more detailed picture of the market, and as a result, the performance of the market improves. The provider $1$ offers a higher price compared to the other providers and attracts mostly business users, while its share of low-profile users drops. On the contrary, the providers $3$ and $4$ offer low prices and attract the largest percentage of low-profile users, while their share of business users becomes low. The strong providers (i.e., the ones with the largest cellular capacity) focus on users with high willingness to pay and QoS requirements, while the weak providers focus on users with low willingness to pay and QoS requirements. This improves the overall performance of users (Fig. \ref{corr_disc}) and reduces the intensity of competition allowing for a higher revenue for providers (Figs. \ref{corr_provider_1} and \ref{corr_provider_4}). In general, as the number of clusters increases, the performance of the market improves. However, the computational complexity of computing the equilibriums of users and providers increases (e.g., a non-linear increase in the execution time as shown in Fig. \ref{fig:execution})\footnote{It is in the interest of a provider to select the appropriate number of clusters that results in high revenue and requires a relatively low execution time, if there are time constraints (e.g., $9$ clusters).}.  

Under high traffic demand, a large decrease of the benefits of providers is observed from a certain point onwards (Figs. \ref{corr_provider_1} and \ref{corr_provider_4}). At this point, the capacity of the networks of providers is reached and disconnected users start appearing. The first users that become disconnected are the ones with a low willingness to pay. Providers enter a competition and restrain their prices aiming to prevent those users from becoming disconnected. Even the provider $1$ enters the competition regardless of its focus on business users. As the traffic demand keeps increasing, it becomes less beneficial for providers to keep those users in the market. This weakens the competition allowing for a small recovery of the revenue gains of providers.  

{\em Case b:} If the maximum willingness to pay ($w^j_R$) and tolerance on low data rate ($h_j$) are independent, the modeling of users at a finer level of detail is beneficial for users: The percentage of disconnected users is significantly reduced (Fig. \ref{ind_disc}). Interestingly, the performance of providers is not always improved. Under a low traffic demand, providers achieve significant revenue gains, while under large traffic demand, they lose revenue compared to a macroscopic modeling of users (Figs. \ref{ind_provider_1} and \ref{ind_provider_4}). Furthermore, the larger the number of clusters, the more prominent the losses. This is a counter-intuitive result: One would expect that the larger the degree of knowledge about the user population, the more significant the benefits of the providers. To explain this phenomenon, we should focus on the distribution of the user groups in this scenario (shown in Fig. \ref{fig:User_population}).  

We distinguish four different types of groups: the business users, low-profile users, lenient users, and value-for-money users. The revenue losses are mainly due to the pressure that value-for-money users and lenient users introduce in the market. Value-for-money users have a low willingness to pay (i.e., low $w^j_R$) and small tolerance on low data rate (i.e., low $h_j$). Satisfying the requirements of those users can be difficult. Providers should offer services of high data rate on low prices. Under a large traffic demand, when the capacity of the networks of providers is reached, value-for-money users start becoming disconnected. When modeling users at high levels of detail, providers are aware of the presence of value-for-money users and restrict their prices in an effort to attract these users given that they correspond to a significant percentage of the market (around 25\%). This results in a steep decrease of the provider revenues (Figs. \ref{ind_provider_1} and \ref{ind_provider_4}). As the number of clusters increases, providers become aware of more ``extreme'' cases of value-for-money users with stricter price and data rate requirements. Therefore, they become more aggressive in the decrease of their prices, and as a result, they lose more revenue. 

%
%

%
\begin{figure*}[t!] 
  \centering
\subfloat[]{\label{prices_corr}\includegraphics[width=0.35\textwidth, height=42mm]{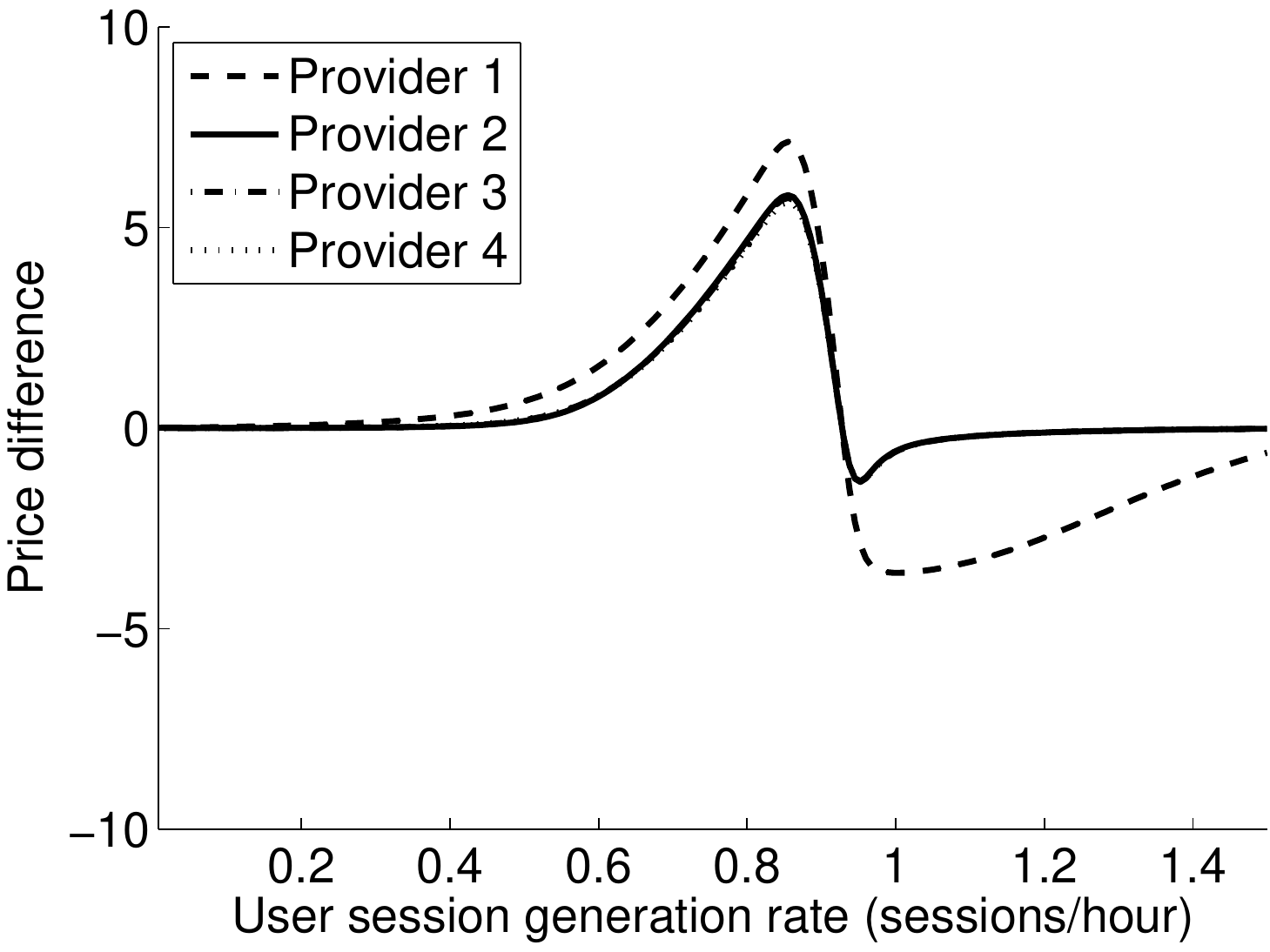}}
\subfloat[]{\label{share_corr}\includegraphics[width=0.35\textwidth, height=42mm]{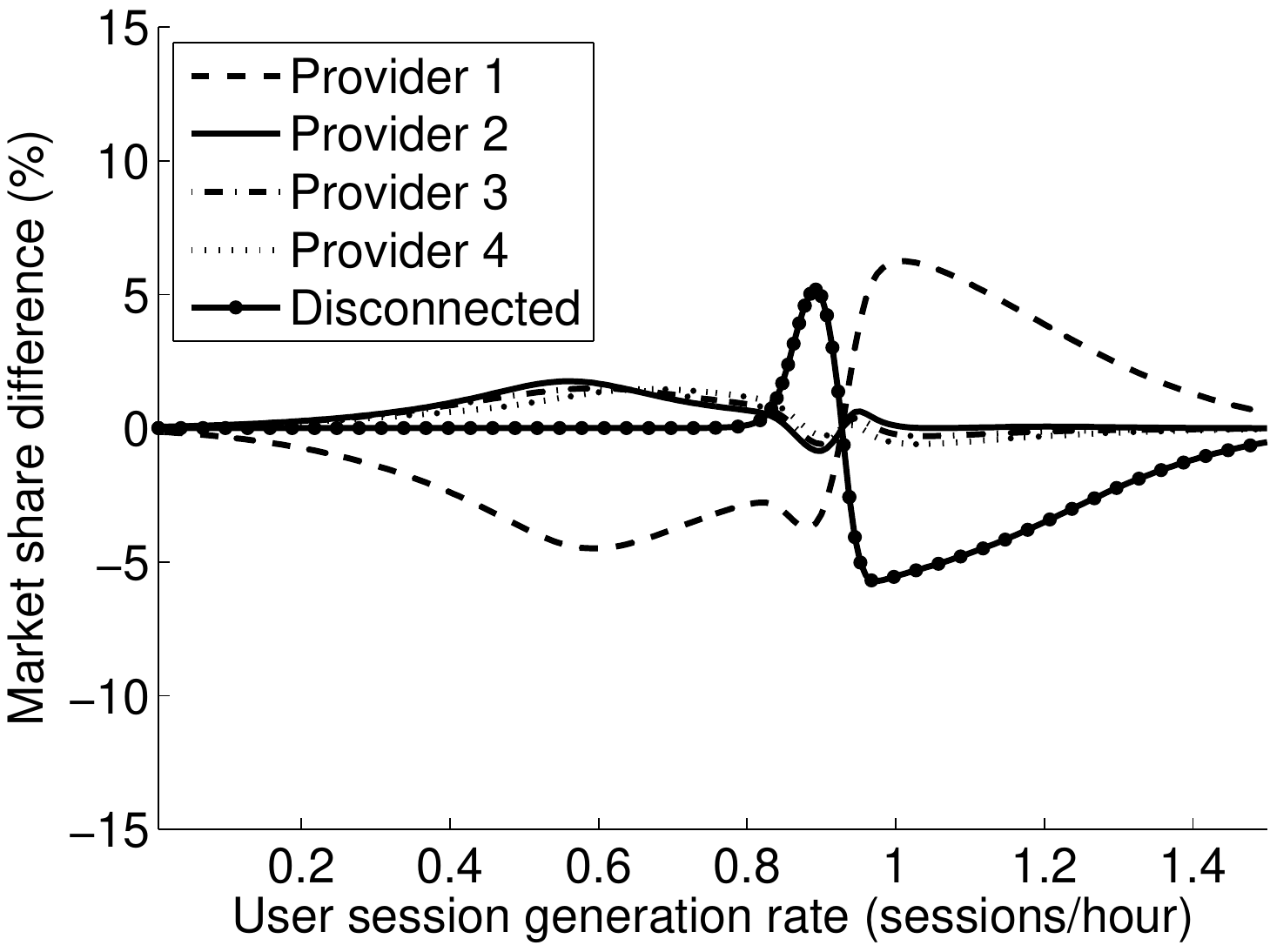}}
\subfloat[]{\label{rev_corr}\includegraphics[width=0.35\textwidth, height=42mm]{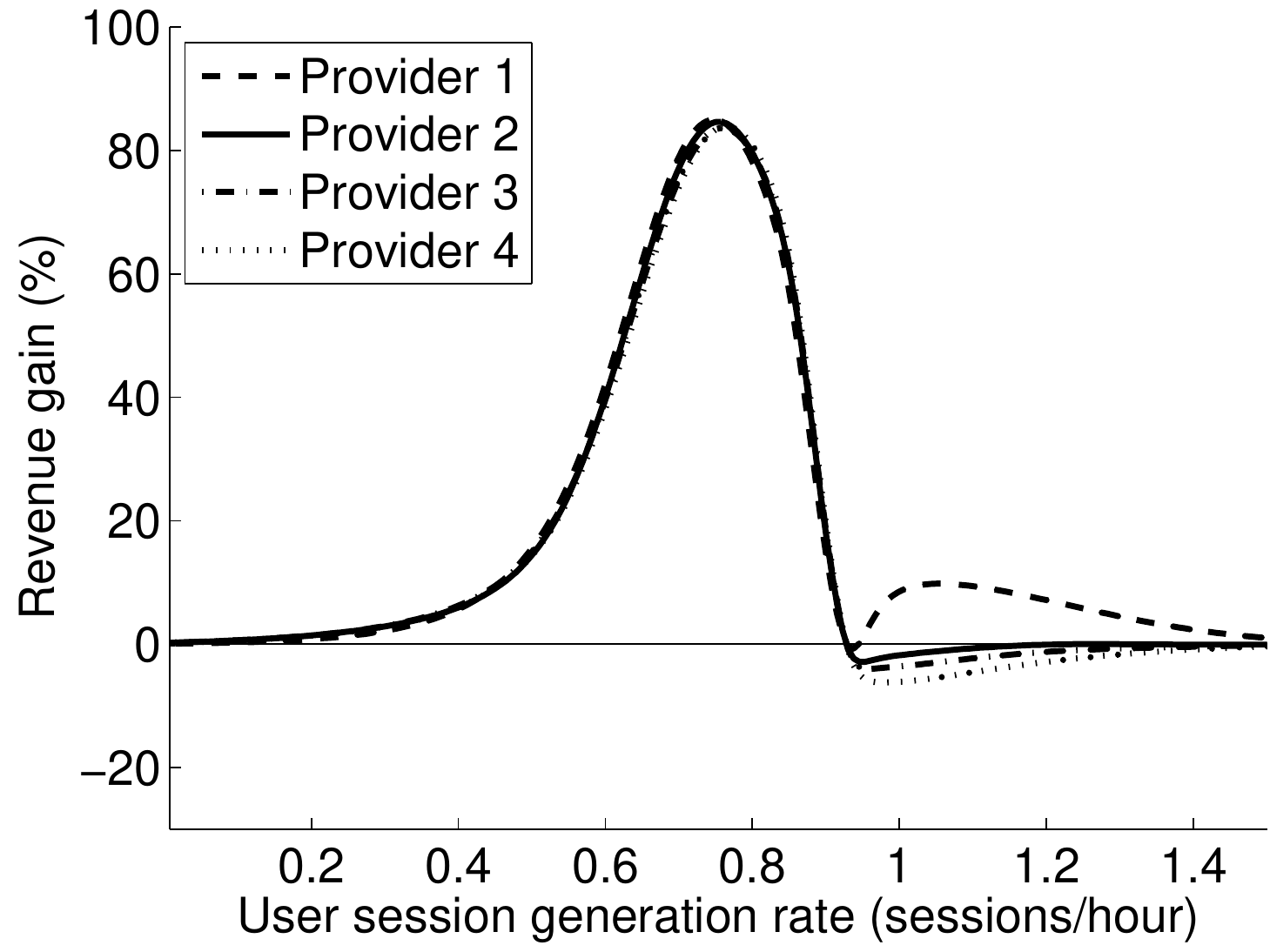}}\\
\subfloat[]{\label{prices_inde}\includegraphics[width=0.35\textwidth, height=42mm]{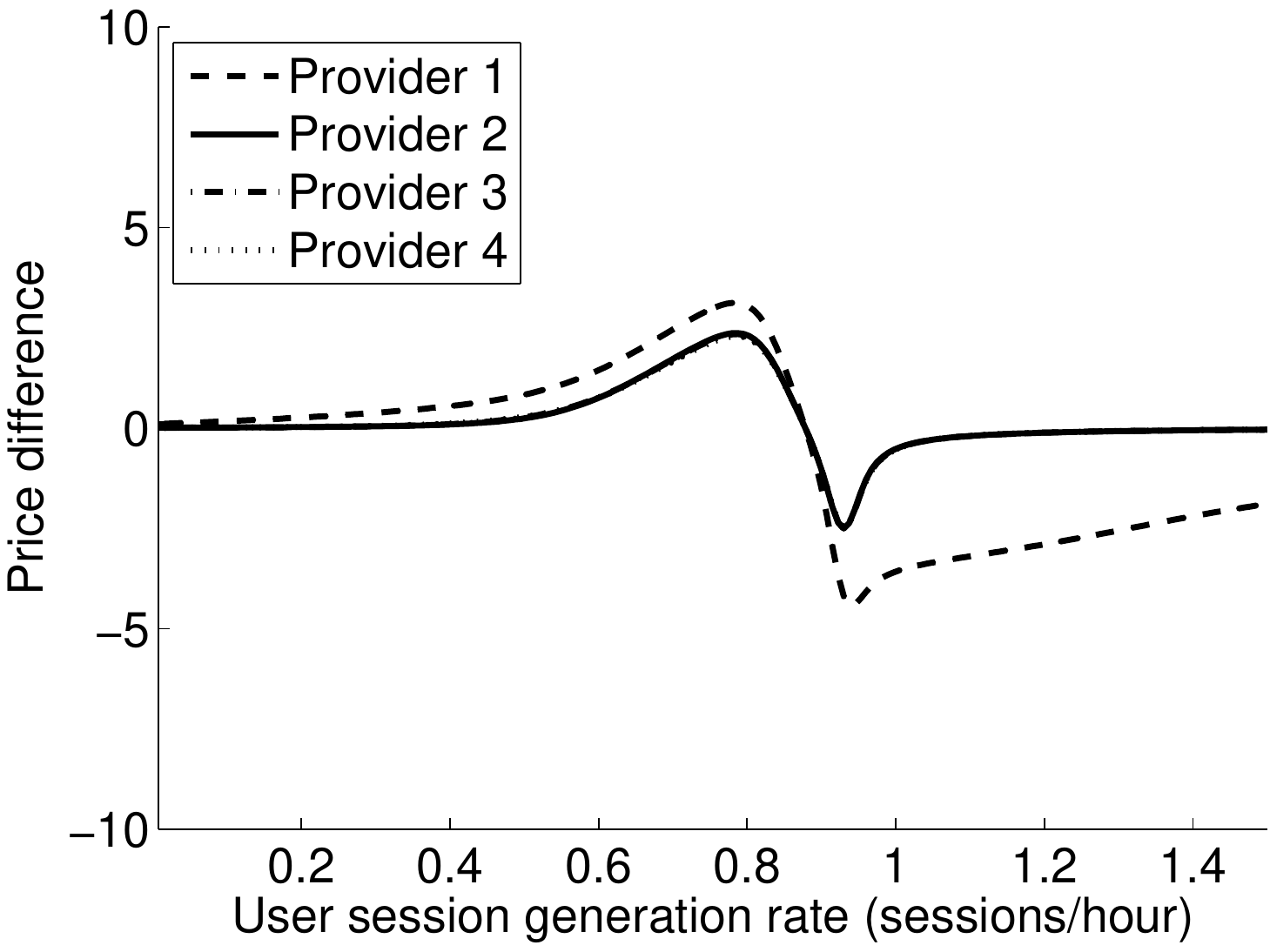}}
\subfloat[]{\label{share_inde}\includegraphics[width=0.35\textwidth, height=42mm]{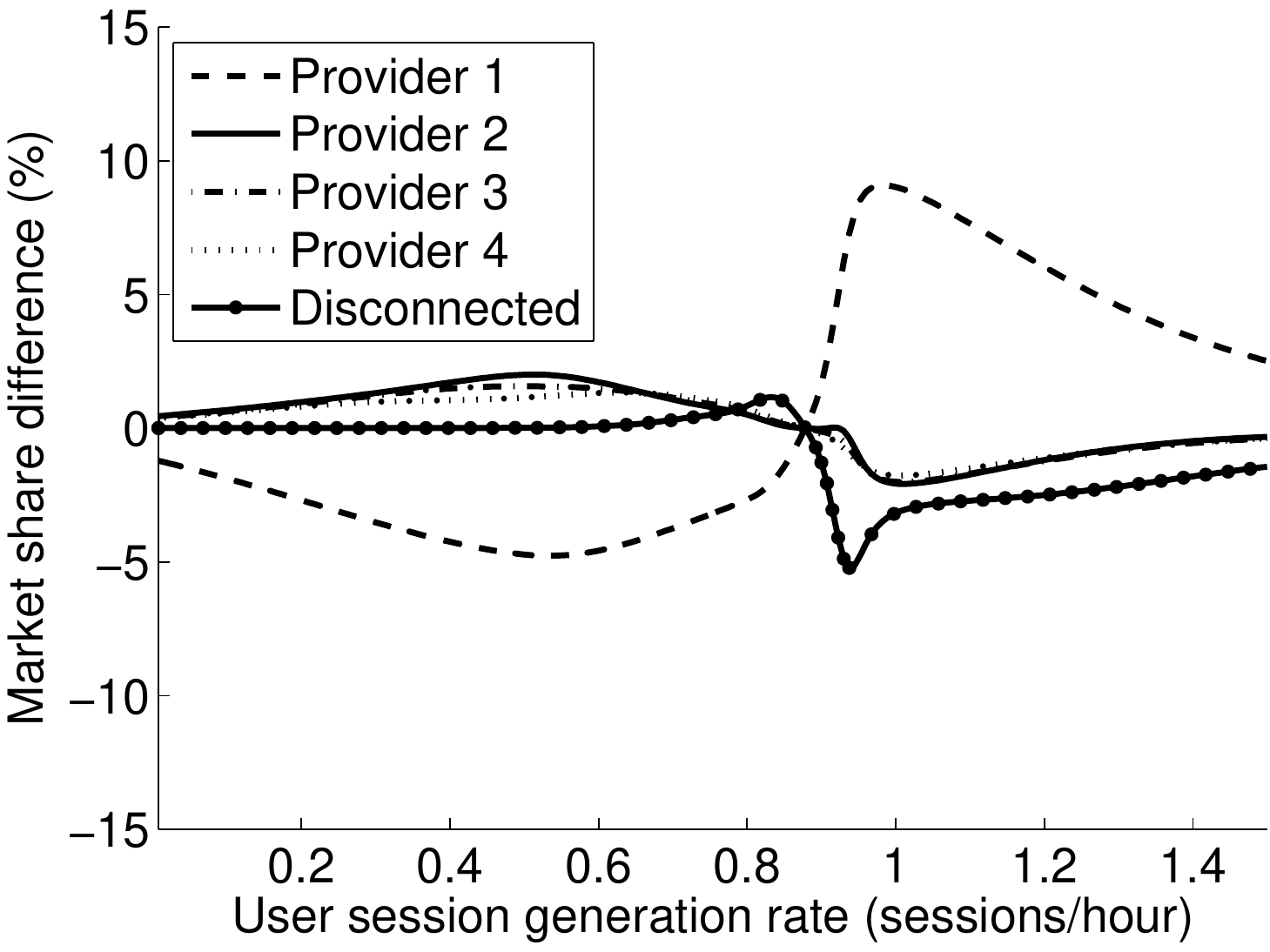}}
\subfloat[]{\label{rev_inde}\includegraphics[width=0.35\textwidth, height=42mm]{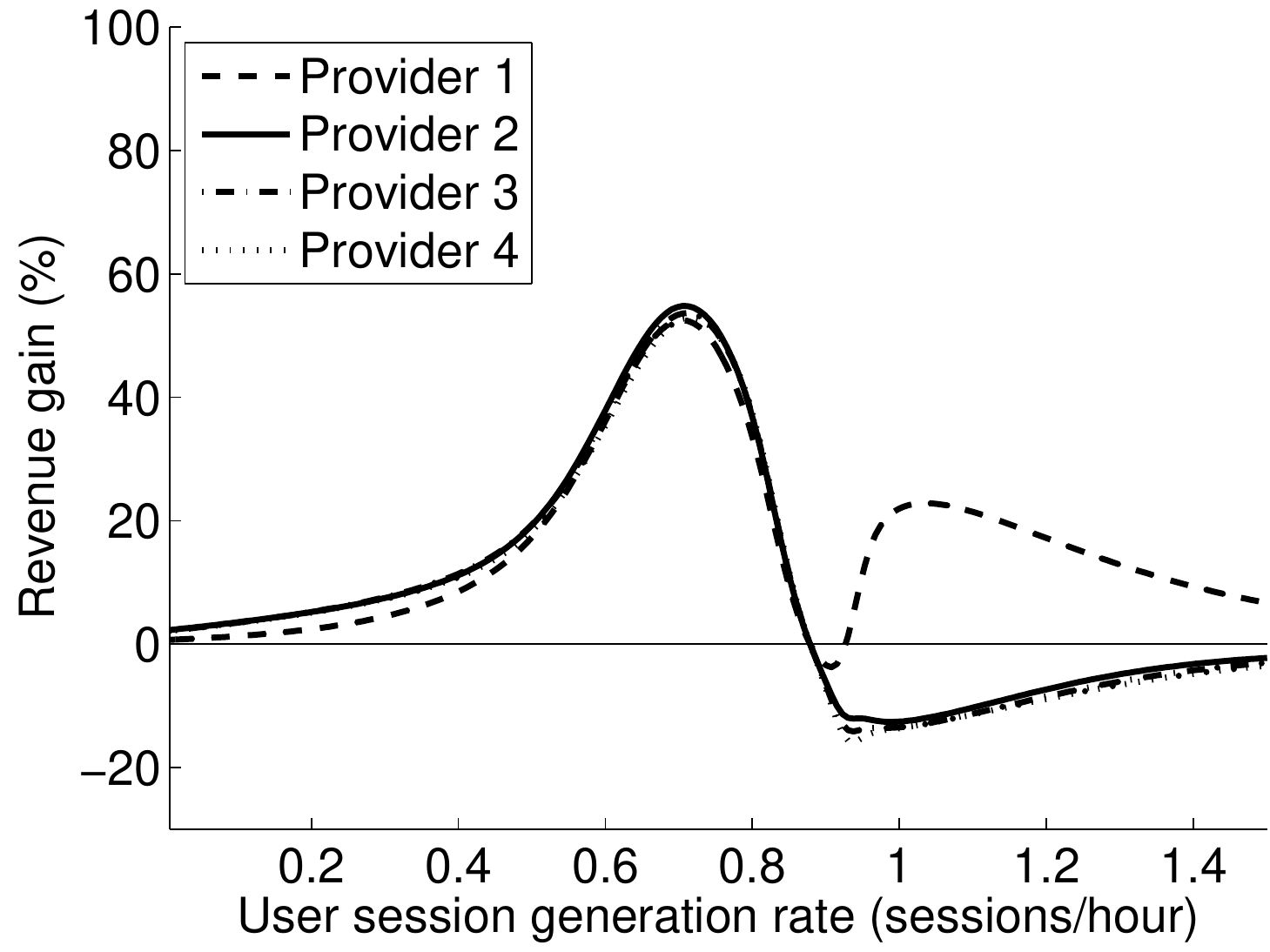}}
  \caption{Performance gains of a market in which only the provider $1$ models users with $9$ clusters, while all other providers model users macroscopically compared to a market in which all providers model users macroscopically. The top (bottom) figures correspond to a scenario in which $w_R^j$ and $h_j$ are correlated (independent), respectively.}
  \label{fig:different_knowledge}
\end{figure*}

In the case of high traffic demand, value-for-money users eventually become disconnected and their influence weakens allowing for a slow recovery of the provider revenues. However, in the case of the provider $4$, its revenue gain always remains negative under large traffic demand (Fig. \ref{ind_provider_4}). This is due to the influence of lenient users. These users have a high willingness to pay (high $w^j_R$) and high tolerance on low data rate (high $h_j$). Price is the parameter that mainly drives their decisions. Under a large traffic demand, disconnected users appear in the value-for-money, low-profile and business users. However, almost all lenient users remain connected due to their low data rate requirements and high willingness to pay. Therefore, as the traffic demand increases, the influence of lenient users in the market intensifies. This strengthens the competition of providers keeping their revenues low. 

{\em Impact of different degrees of detail in the knowledge about customers among providers}. We repeated the analysis, considering now a market in which only the provider $1$ models users with $9$ clusters, while all other providers model users macroscopically. Figs. \ref{prices_corr}, \ref{share_corr}, and \ref{rev_corr} show the differences in the prices, market share, and revenue of providers, respectively, compared to a market in which all providers model users macroscopically when $w_R^j$ and $h_j$ are correlated. Figs. \ref{prices_inde}, \ref{share_inde}, and \ref{rev_inde} present the same differences when $w^j_R$ and $h_j$ are independent.

The provider $1$ always achieves revenue gains. This is observed both when $w_R^j$ and $h_j$ are correlated and when they are independent (Figs. \ref{rev_corr} and \ref{rev_inde}, respectively). With its higher degree of knowledge, the provider $1$ can influence the market to its benefit outsmarting the other providers. This is an important result which shows that the benefits of a provider from modeling the user population in a high level of detail strongly depend on the level of knowledge of the other providers about users. 

Another interesting trend is that the effect of the knowledge of the provider $1$ on the revenues of the other providers is twofold. Under small traffic demand, the provider $1$ raises its price compared to the one offered at the macroscopic level (Figs. \ref{prices_corr} and \ref{prices_inde}). Given that it is the most influential provider in the market, its price increase provides also an opportunity to the other providers to raise their prices. This results in significant revenue gains for all providers (Figs. \ref{rev_corr} and \ref{rev_inde}). However, when the traffic demand becomes large, suddenly, the provider $1$ ``turns against'' the other providers. When the capacity of the networks of providers is reached and disconnected users appear, it reduces its price bellow the prices of the other providers to attract those users (Figs. \ref{prices_corr} and \ref{prices_inde}). Given their macroscopic view of the market, the other providers do not react. This results in an increase of the revenue gain of the provider $1$ at the expense of the revenues of the other providers (Figs. \ref{rev_corr} and \ref{rev_inde}).
\begin{figure*}[t!] 
  \centering
\subfloat[]{\label{prices_corr_prov_4}\includegraphics[width=0.35\textwidth, height=42mm]{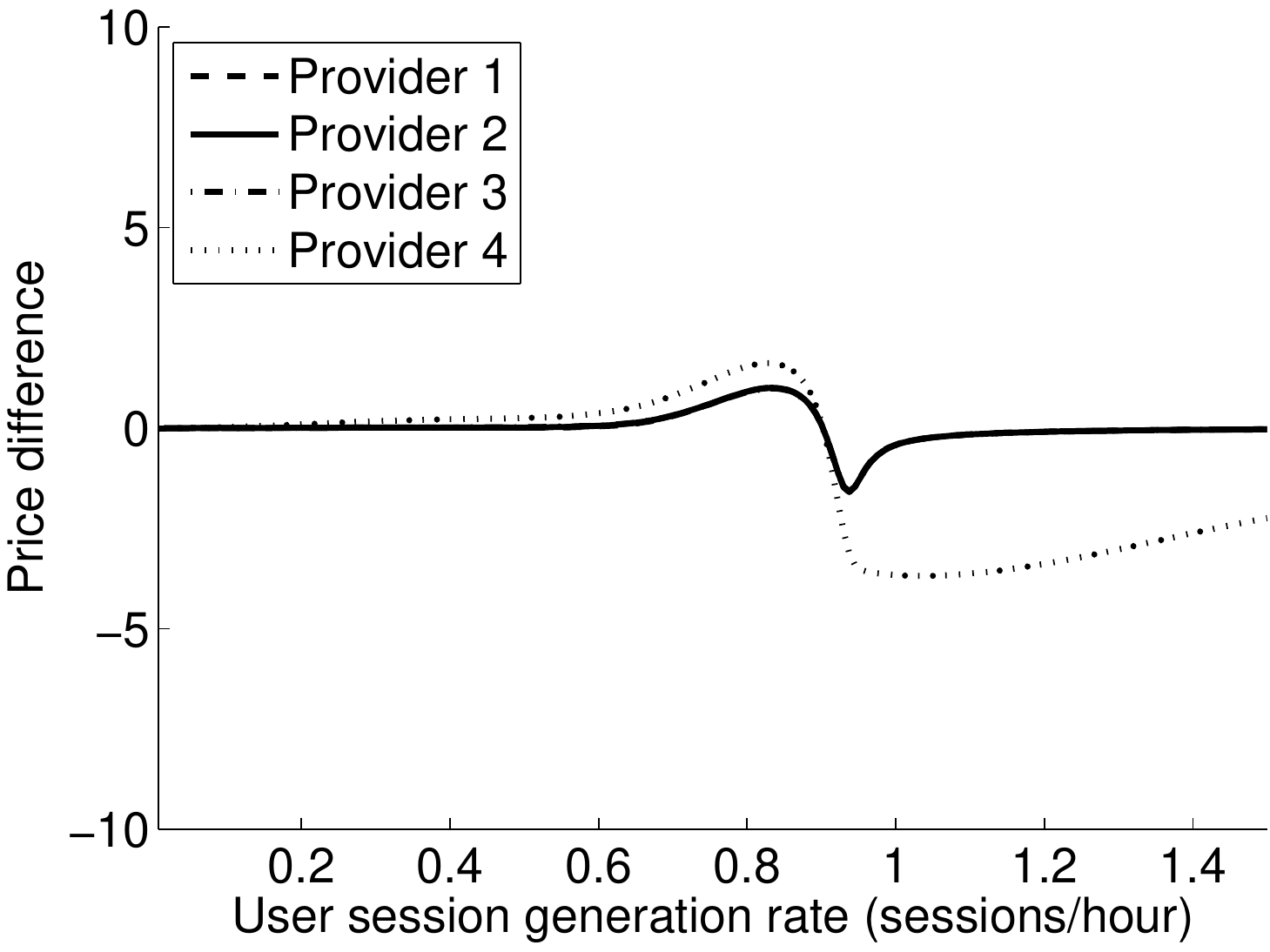}}
\subfloat[]{\label{share_corr_prov_4}\includegraphics[width=0.35\textwidth, height=42mm]{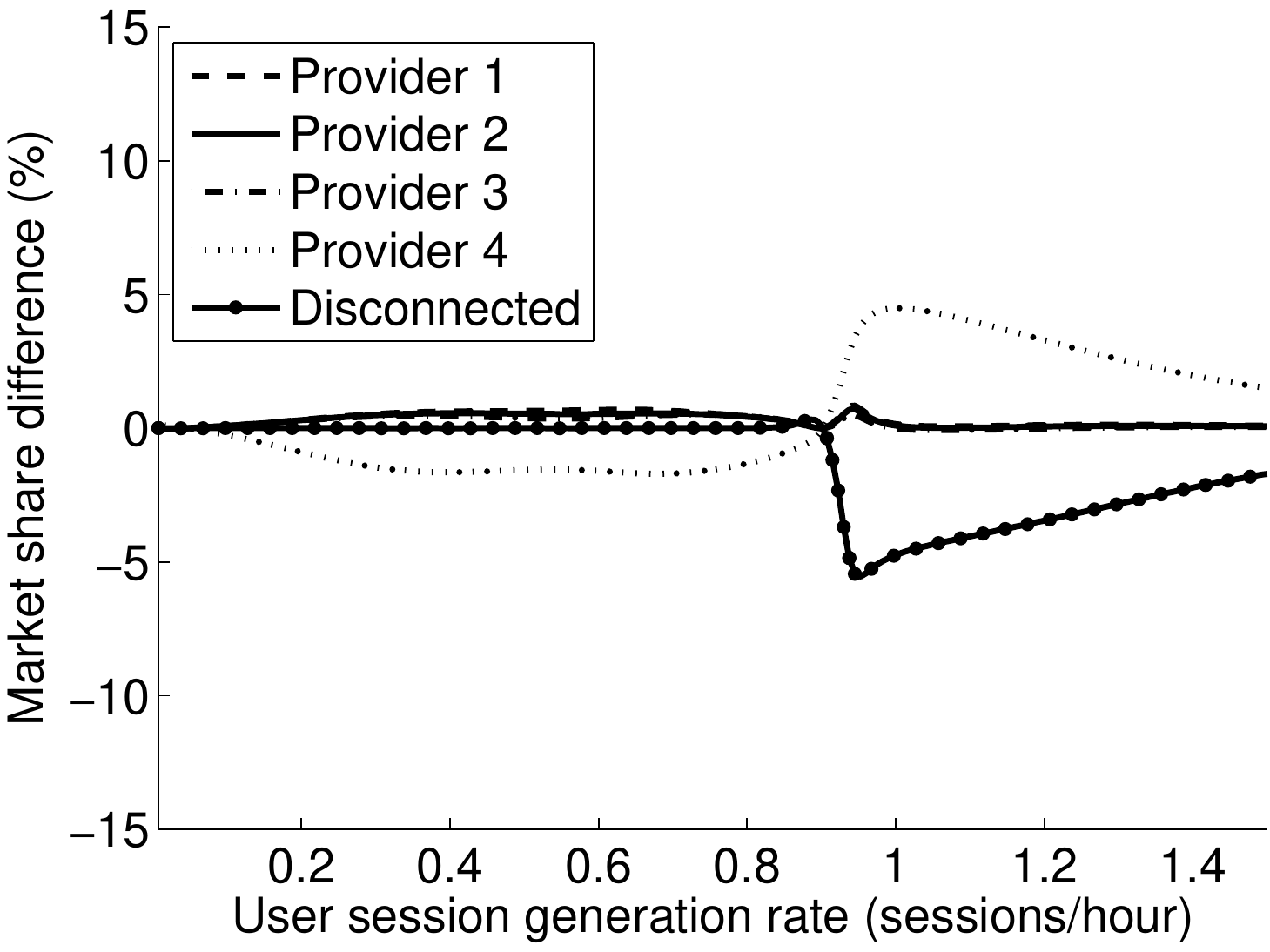}}
\subfloat[]{\label{rev_corr_prov_4}\includegraphics[width=0.35\textwidth, height=42mm]{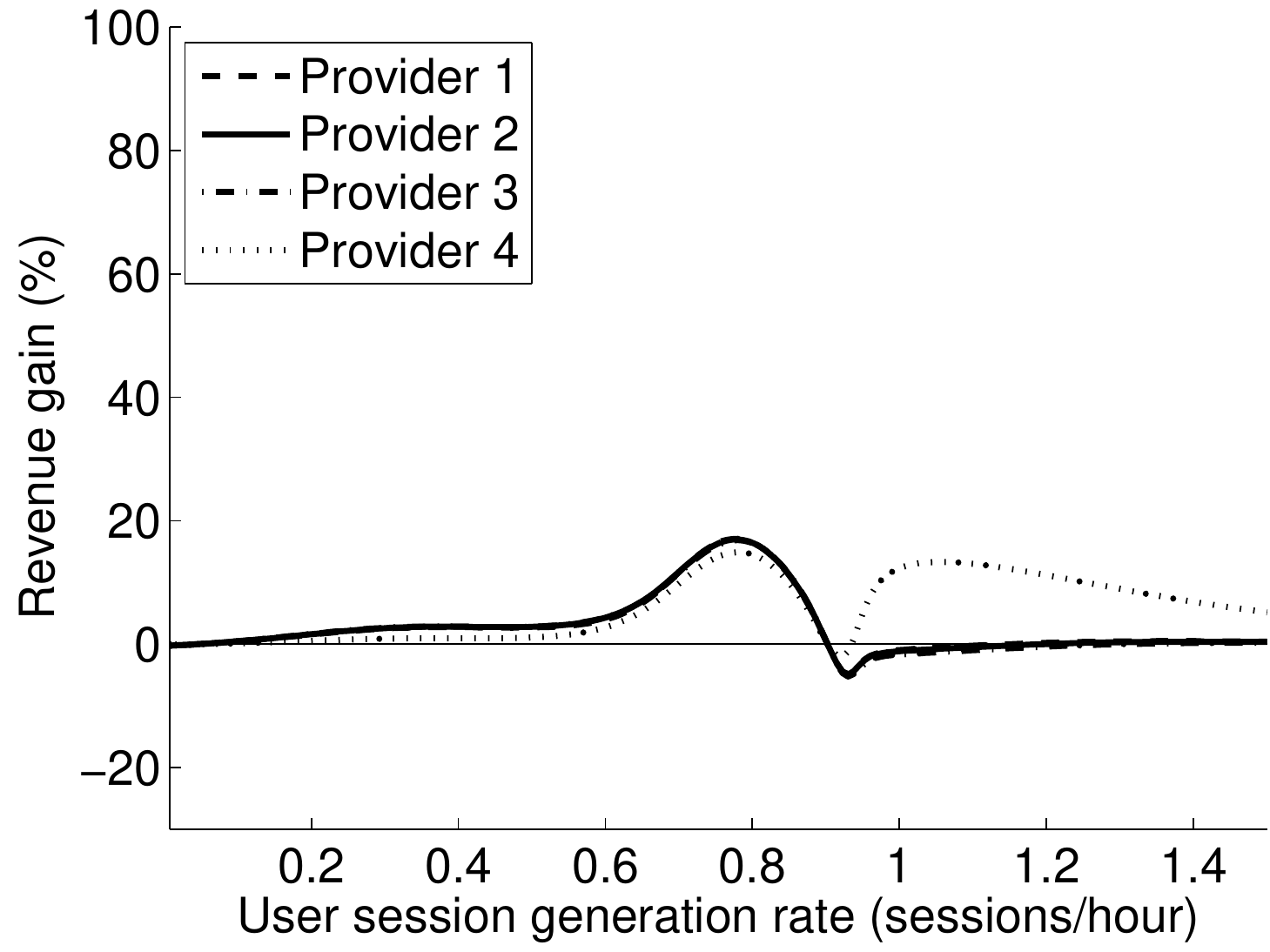}}\\
\subfloat[]{\label{prices_inde_prov_4}\includegraphics[width=0.35\textwidth, height=42mm]{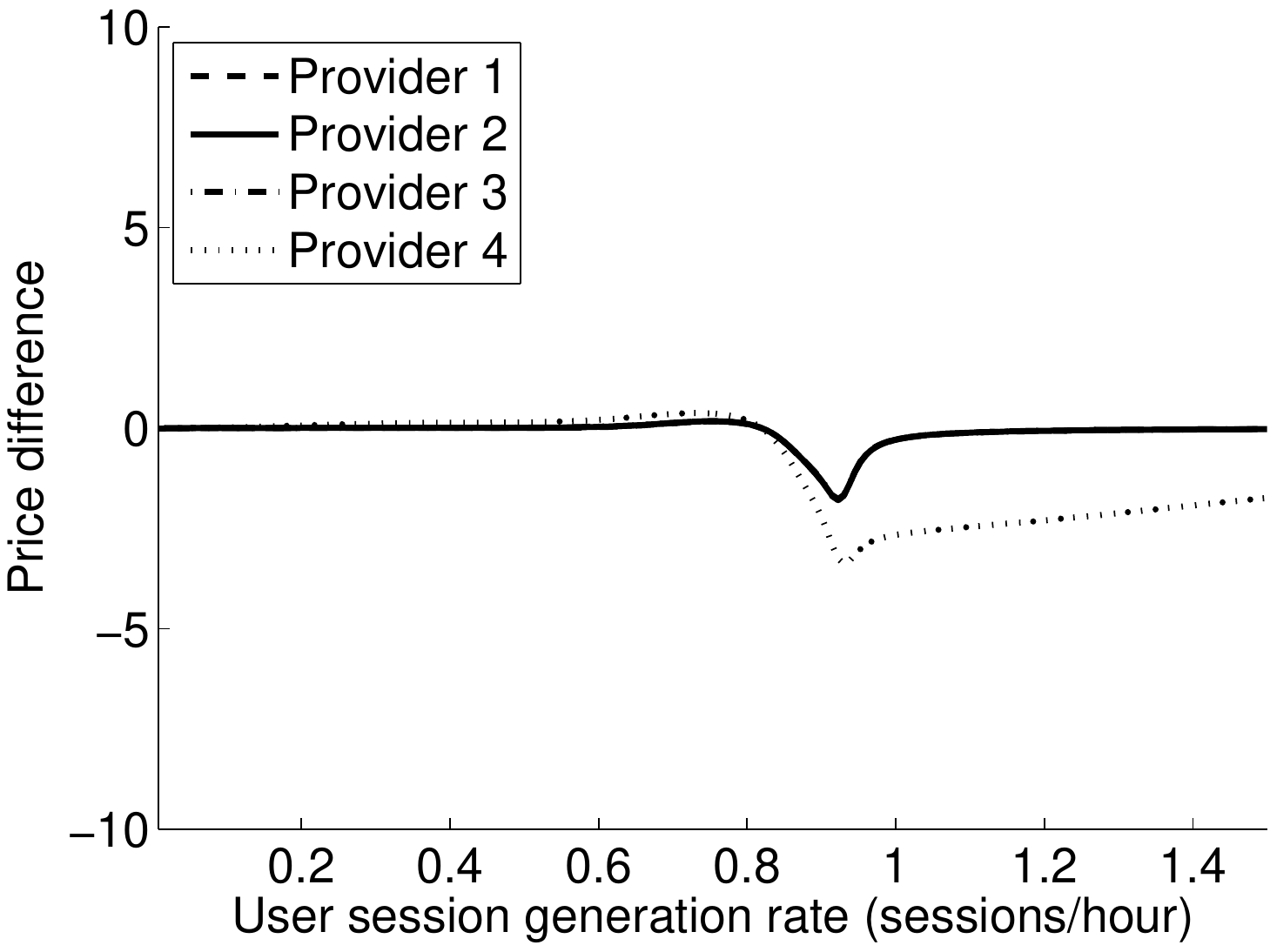}}
\subfloat[]{\label{share_inde_prov_4}\includegraphics[width=0.35\textwidth, height=42mm]{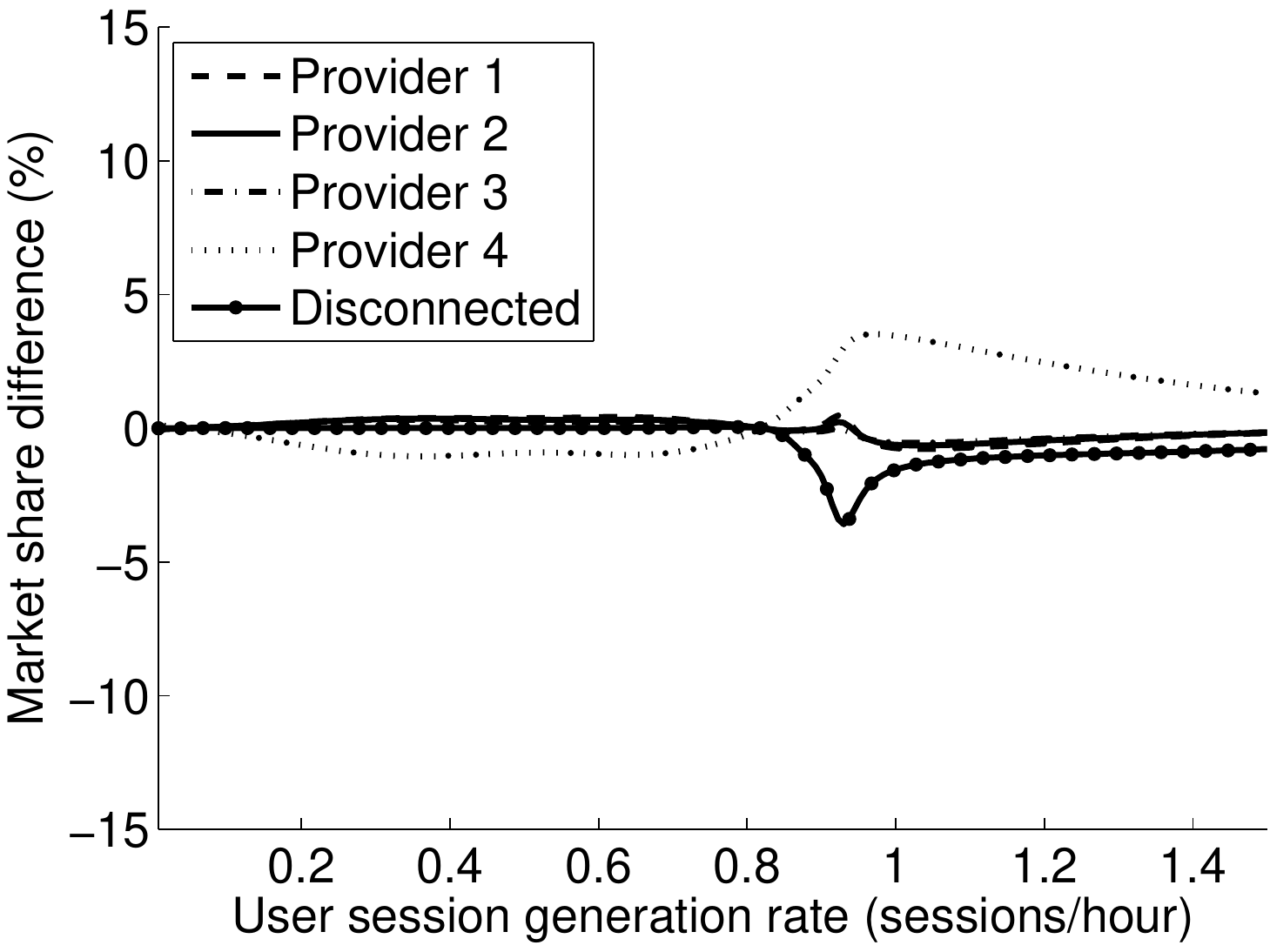}}
\subfloat[]{\label{rev_inde_prov_4}\includegraphics[width=0.35\textwidth, height=42mm]{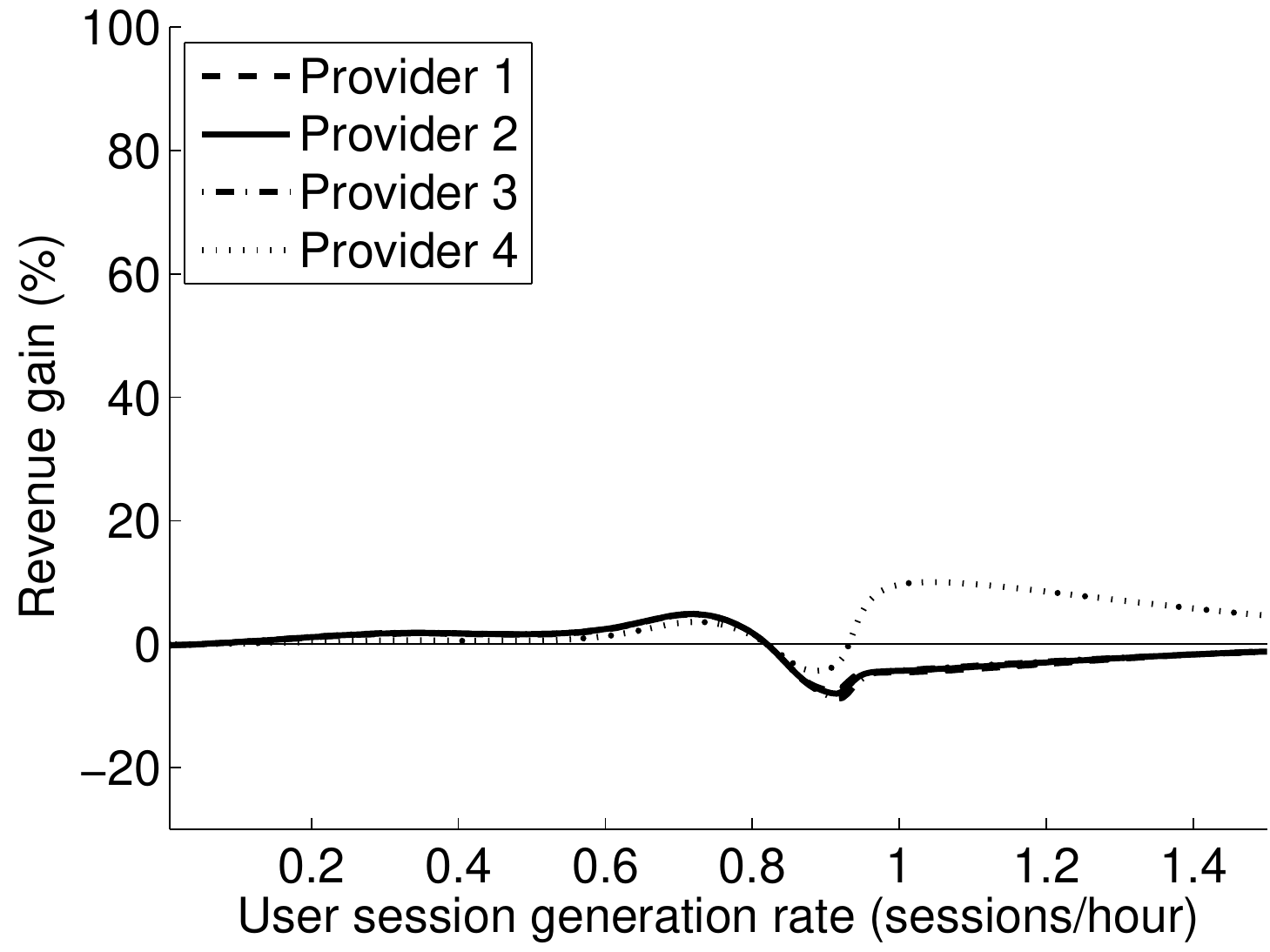}}
  \caption{Performance gains of a market in which only the provider $4$ models users with $9$ clusters, while all other providers model users macroscopically compared to a market in which all providers model users macroscopically. The top (bottom) figures correspond to a scenario in which $w_R^j$ and $h_j$ are correlated (independent), respectively.}
  \label{fig:different_knowledge_provider_4}
\end{figure*}

Let us now explain the behaviour of the provider $1$ in more detail. When the user traffic demand is low, the provider $1$ realizes that it is more beneficial to focus on business users increasing its price. This results in a decrease of its market share compared to the one obtained at the macroscopic level (Figs. \ref{share_corr} and \ref{share_inde}). However, this decrease of market share is useful because it improves the quality of service of the provider $1$ making its subscription more appealing to business users. Given their improved satisfaction, those users can pay a higher price to the provider $1$ increasing its revenue. This trend persists as the traffic demand increases until the capacity of the networks of providers is reached. From this point onwards, the quality of service drops substantially and some of the business users become dissatisfied and decide to disconnect. Then, the provider $1$ suddenly changes its strategy. Instead of focusing on the business users, it realizes that it is more beneficial to restrict its price in order to prevent users with low willingness to pay from becoming disconnected and attract them to its network. Other providers do not realize this early enough due to their macroscopic view of the market and they do not react. That way, the provider $1$ achieves significant revenue gain at the expense of the revenues of the other providers. 

We repeated the previous analysis considering now that only the provider $4$ models users with $9$ clusters, while all other providers model users macroscopically (Fig. \ref{fig:different_knowledge_provider_4}). The provider $4$ achieves revenue gains both when $w^j_R$ and $h_j$ are correlated (Fig. \ref{rev_corr_prov_4}) and when they are not (Fig. \ref{rev_inde_prov_4}). Additionally, when the traffic demand is low, all providers achieve revenue gains, while under large traffic only the provider $4$ gains additional revenue at the expense of the revenues of the other providers (Figs. \ref{rev_corr_prov_4} and \ref{rev_inde_prov_4}). However, given that the provider $4$ is the weakest one in the market, its influence is low, and therefore, the observed tends of Fig. \ref{fig:different_knowledge_provider_4} are subtler compared to Fig. \ref{fig:different_knowledge}. 

\subsection{Offering of multiple dataplans} \label{sec:multiple_services}
\begin{figure*}[t!] 
\centering
\subfloat[]{\label{mserv_corr_disc}\includegraphics[width=0.35\textwidth, height=42mm]{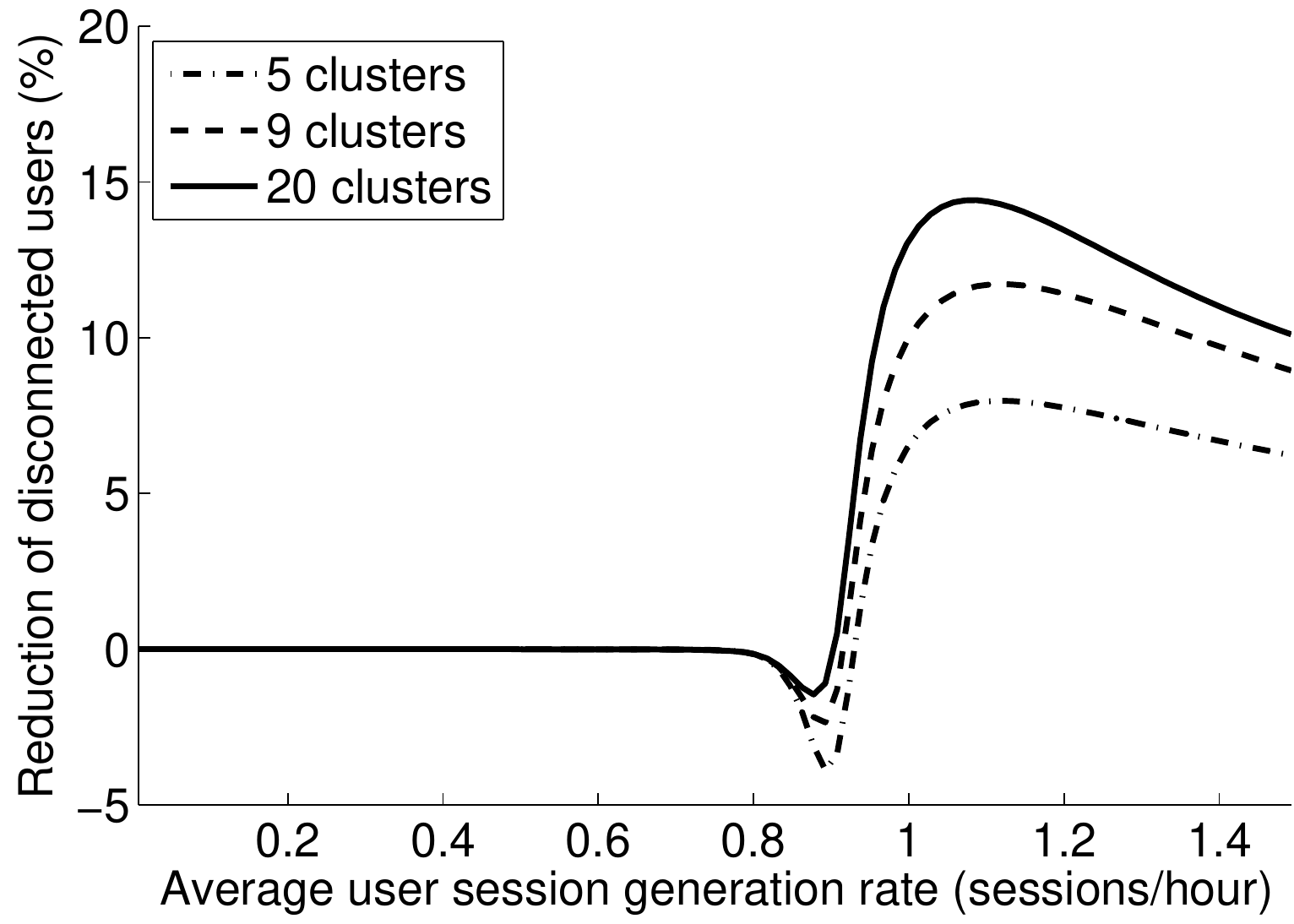}}
\subfloat[]{\label{mserv_corr_provider_1}\includegraphics[width=0.35\textwidth, height=42mm]{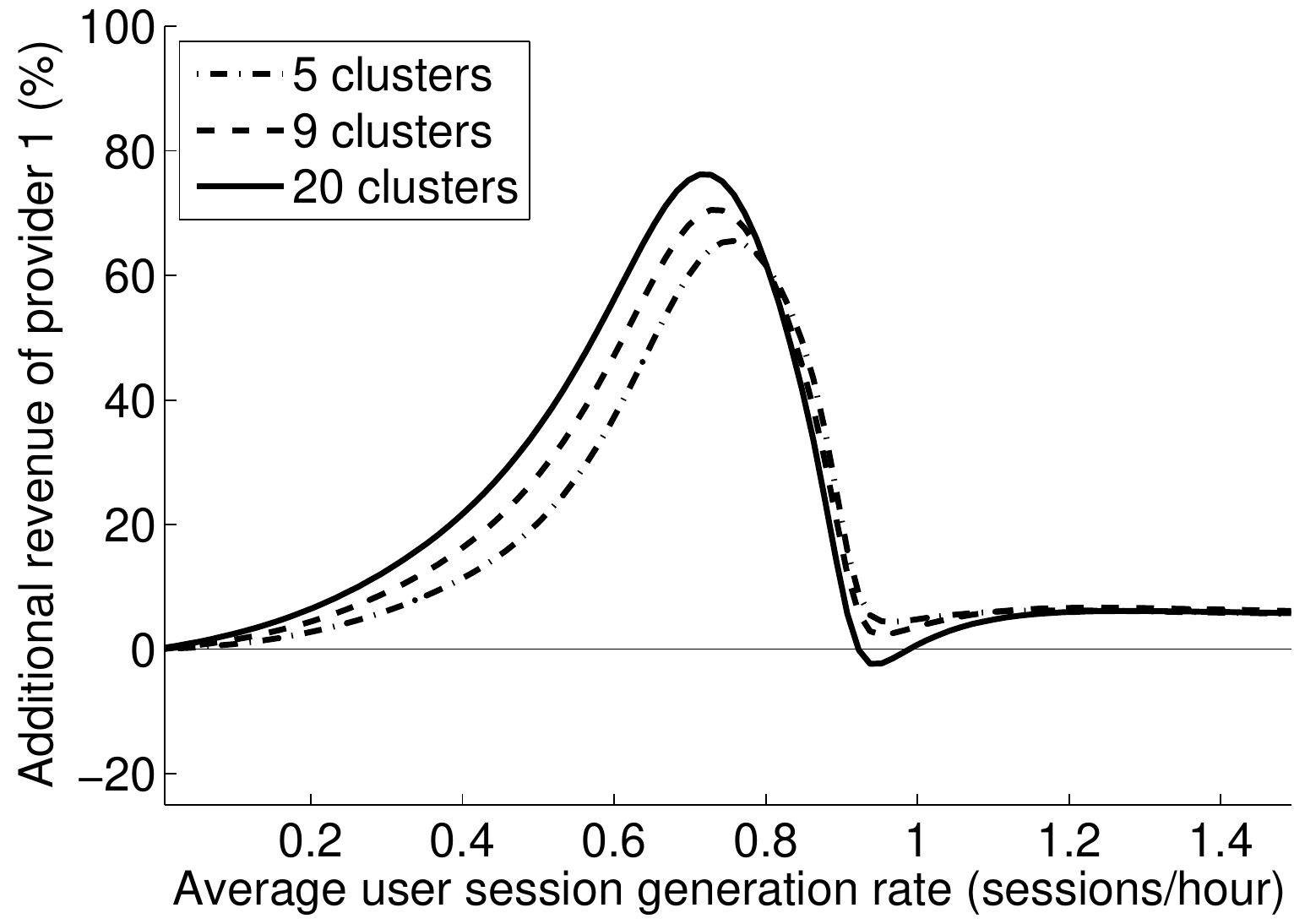}}
\subfloat[]{\label{mserv_corr_provider_4}\includegraphics[width=0.35\textwidth, height=42mm]{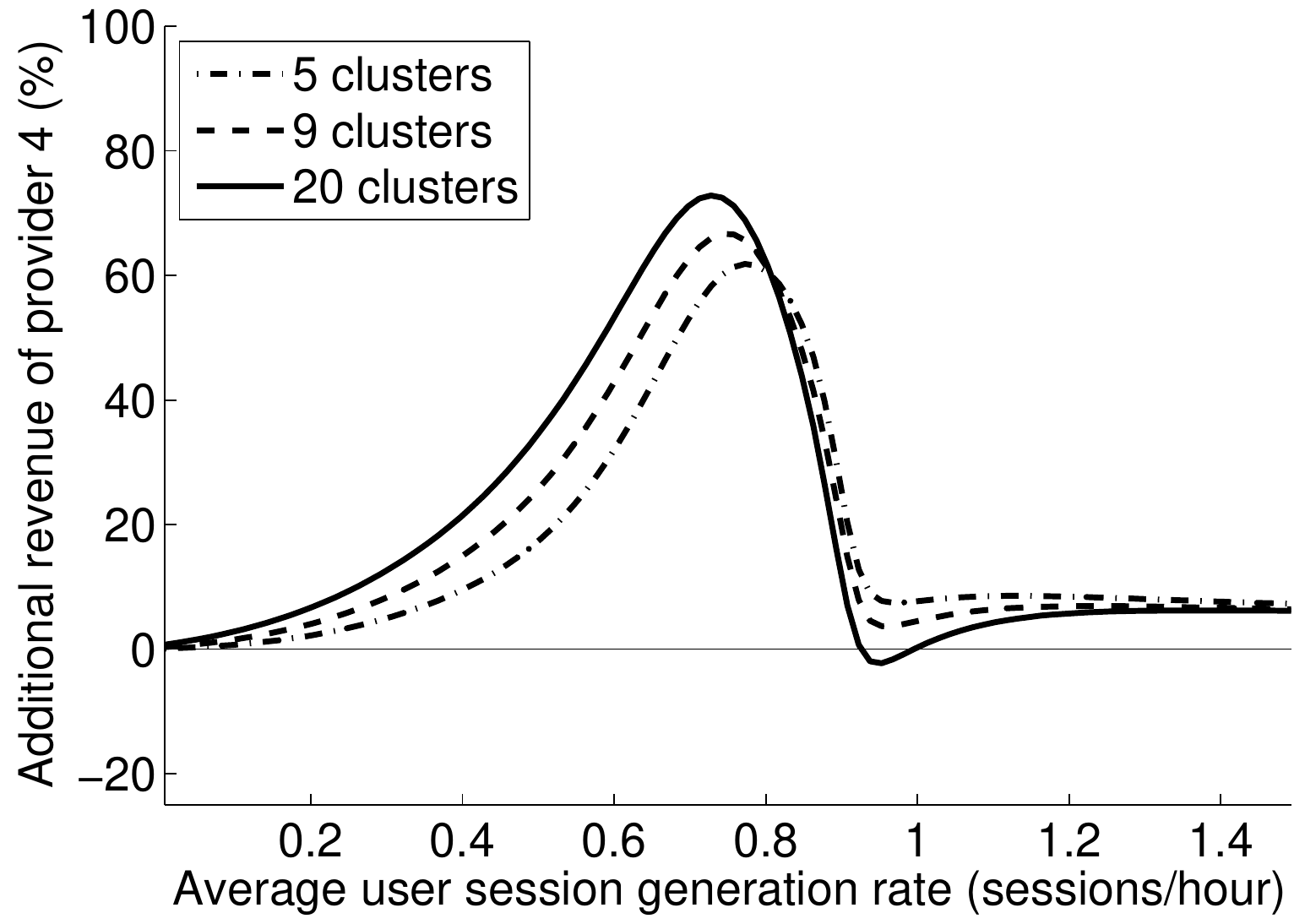}}\\
\subfloat[]{\label{mserv_ind_disc}\includegraphics[width=0.35\textwidth, height=42mm]{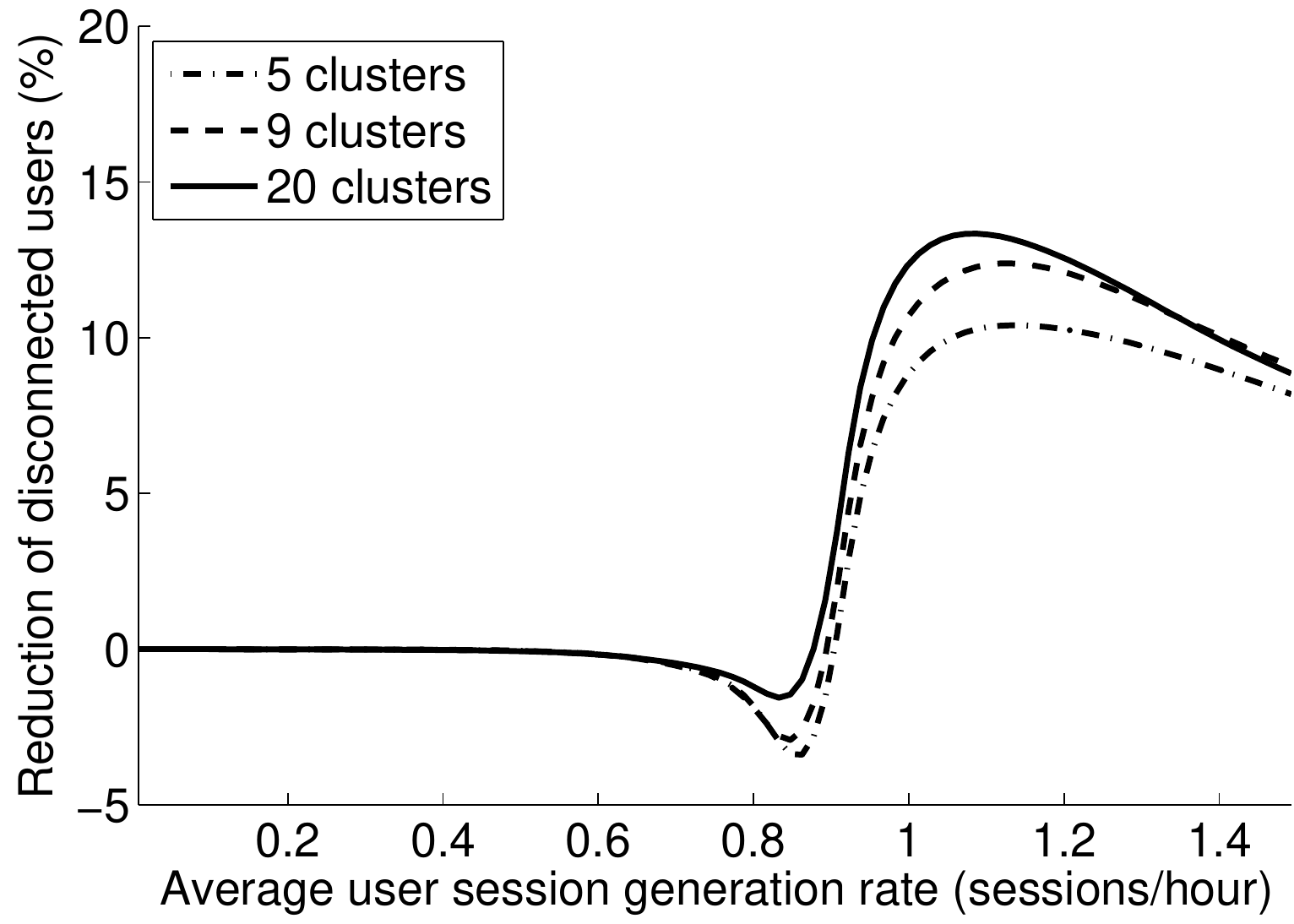}}
\subfloat[]{\label{mserv_ind_provider_1}\includegraphics[width=0.35\textwidth, height=42mm]{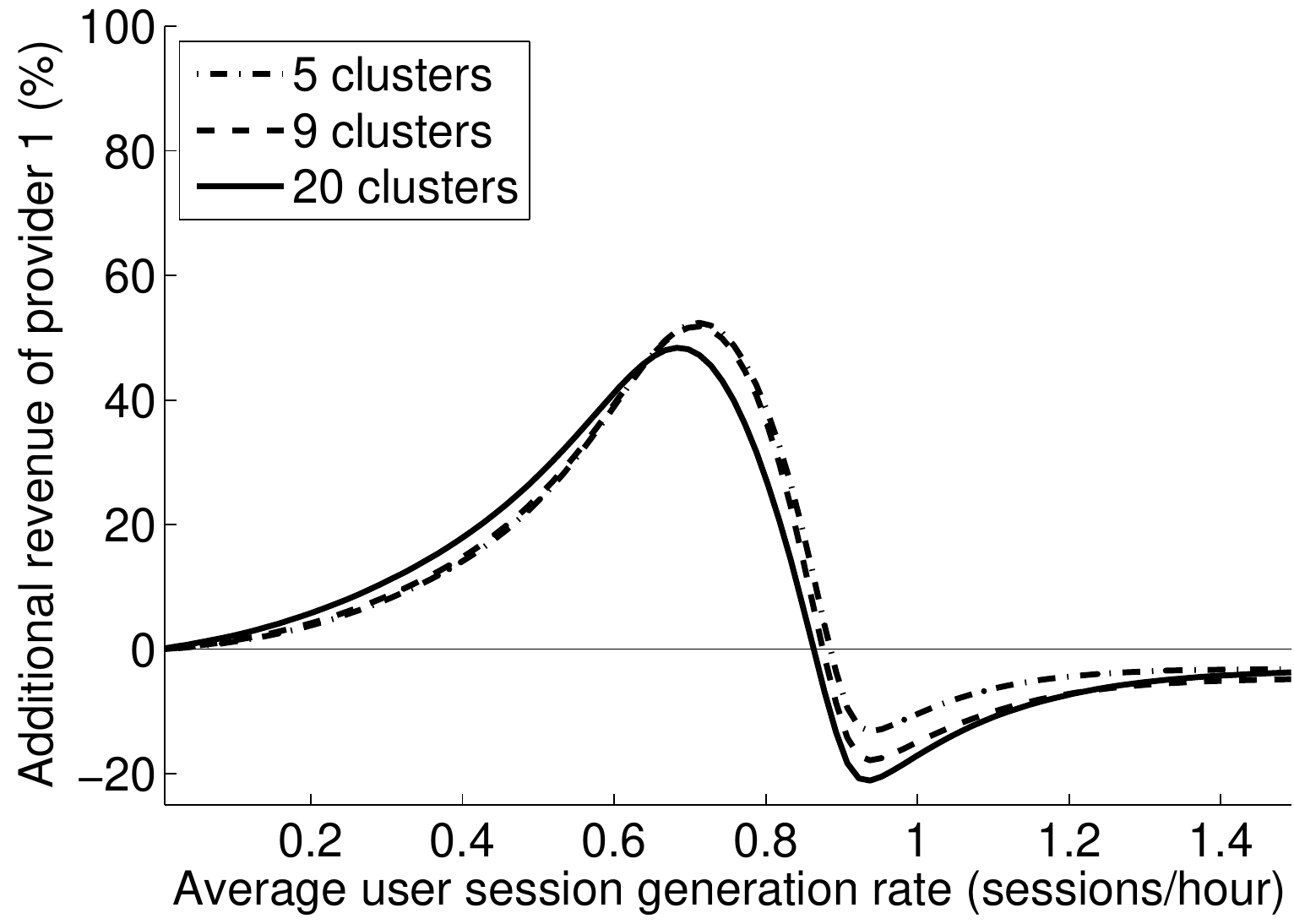}}
\subfloat[]{\label{mserv_ind_provider_4}\includegraphics[width=0.35\textwidth, height=42mm]{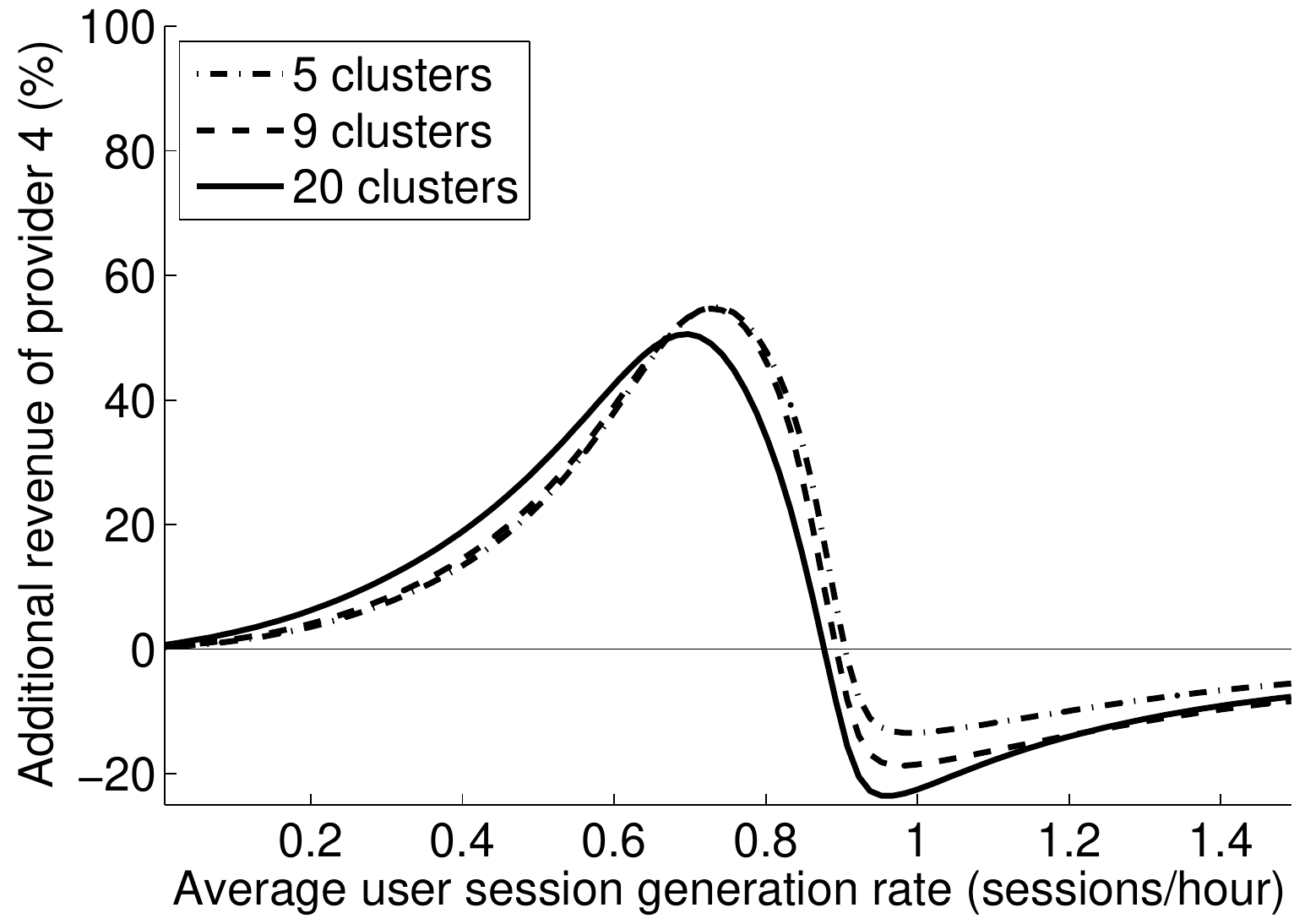}}
\caption{Performance gains when providers model users at different levels of detail and offer $3$ dataplans compared to modeling users macroscopically and offering $1$ dataplan with $w_R^j$, $h_j$, and $n_j$ to be correlated (top) and independent (bottom), respectively.}
\label{fig:mserv}
\end{figure*}

We have defined a market in which users of different groups deviate not only in their willingness to pay ($w^j_R$) and tolerance on low data rate ($h_j$) but also in their traffic demand ($d_j$). 
As in Section \ref{sec:heterogeneity}, $w^j_R$ and $h_j$ follow normal distributions, while the average session generation rate over all user groups ($\overline{\lambda}=\sum_{j=1}^{J} \lambda_j/\sum_{j=1}^{J} N_{j}$) varies from 0 up to 1.5 sessions/hour. The normalized session generation rate $n_j = \frac{\lambda_j/N_j}{\overline{\lambda}}$ of user groups follows a normal distribution of mean 1 and standard deviation $0.2$. Some user groups correspond to ``heavy'' users producing a large amount of traffic (users groups with a large $n_j$), while other groups correspond to ``light'' users (groups with a low $n_j$). Again we distinguish two market scenarios. In the first scenario, the cross correlation of $w_R^j$ and $h_j$ is equal to $-0.85$, while the cross correlation of $w^j_R$ and $n_j$ is $0.85$. This means that users with a large willingness to pay (high value of $w_R^j$) usually are heavy users (high value of $n_j$) and less tolerant on low data rate (low value of $h_j$). In the second scenario, the cross correlation among $w_R^j$, $h_j$, and $n_j$ is equal to $0$ and the maximum willingness to pay, data rate requirements and traffic demand of groups are completely independent. 
In this set of experiments, we set the noise parameter of the Logit dynamics modeling the user-decision making ($\epsilon$) equal to $2$.

We assume that a provider i offers a total number of $S_i$ dataplans each corresponding to a different interval of traffic demand and flat-rate price. The first dataplan is ''addressed" to 
light users with a normalized session generation rate up to $100/S_i$\% percentile, while the last dataplan is appropriate to heavy users with 
normalized session generation rate lying in the interval from $(S_i-1)*100/S_i$\% up to 100\% percentile. 
In other words, the range of the values of the normalized user session generation rate is divided into $S_i$ segments. Users groups belonging to different segments are charged with a different price. 
Providers can also model users at different levels of detail by estimating a different number of user clusters. We have analysed a market in which providers offer $3$ dataplans to users. 
The performance gains obtained when providers model users at different levels of detail compared to when they model users macroscopically and offer only $1$ dataplan are presented in Fig. \ref{fig:mserv}. 
The top (bottom) figures correspond to the case that $w^j_R$, $h_j$, and $n_j$ are correlated (independent), respectively.

Similar trends are observed as in Fig. \ref{fig:multi_layer}. When $w_R^j$, $h_j$, and $n_j$ are correlated, in most cases, providers achieve revenue benefits when they model users at a higher level of detail (Figs. \ref{mserv_corr_provider_1} and \ref{mserv_corr_provider_4}), while the reduction of the percentage of disconnected users becomes more prominent (Fig. \ref{mserv_corr_disc}). However, in a small interval around $0.95$ session/hour, providers lose a small amount of revenue when they model users in higher detail. When the number of user clusters increases, an increase in the number of dataplans is required for a better pricing. With a larger number of dataplans, providers can charge the different user clusters more efficiently achieving higher revenue. In the cases of $9$ clusters and $20$ clusters, when the number of dataplans is increased above $3$, the revenue losses around $0.95$ sessions/hour are reduced. We omit those results due to lack of space.

When $w_R^j$, $h_j$, and $n_j$ are independent, the observed trends are exactly the same as in Fig. \ref{fig:multi_layer}. An increase of the number of user clusters results in a more prominent reduction of the percentage of disconnected users (Fig. \ref{mserv_ind_disc}). Additionally, under a low traffic demand, an increase of the number of clusters always results in revenue benefits for providers. However, under a large traffic demand, modeling users with a large number of clusters results in revenue losses compared to the macroscopic case (Figs. \ref{mserv_ind_provider_1} and \ref{mserv_ind_provider_4}). Again those revenue losses are due to the existence of value-for-money users and lenient users. Those users intensify the competition of providers under a large traffic demand resulting in lower offered prices and revenue. 

\begin{figure*}[t!] 
  \centering
\subfloat[]{\label{dserv_corr_disc}\includegraphics[width=0.35\textwidth, height=42mm]{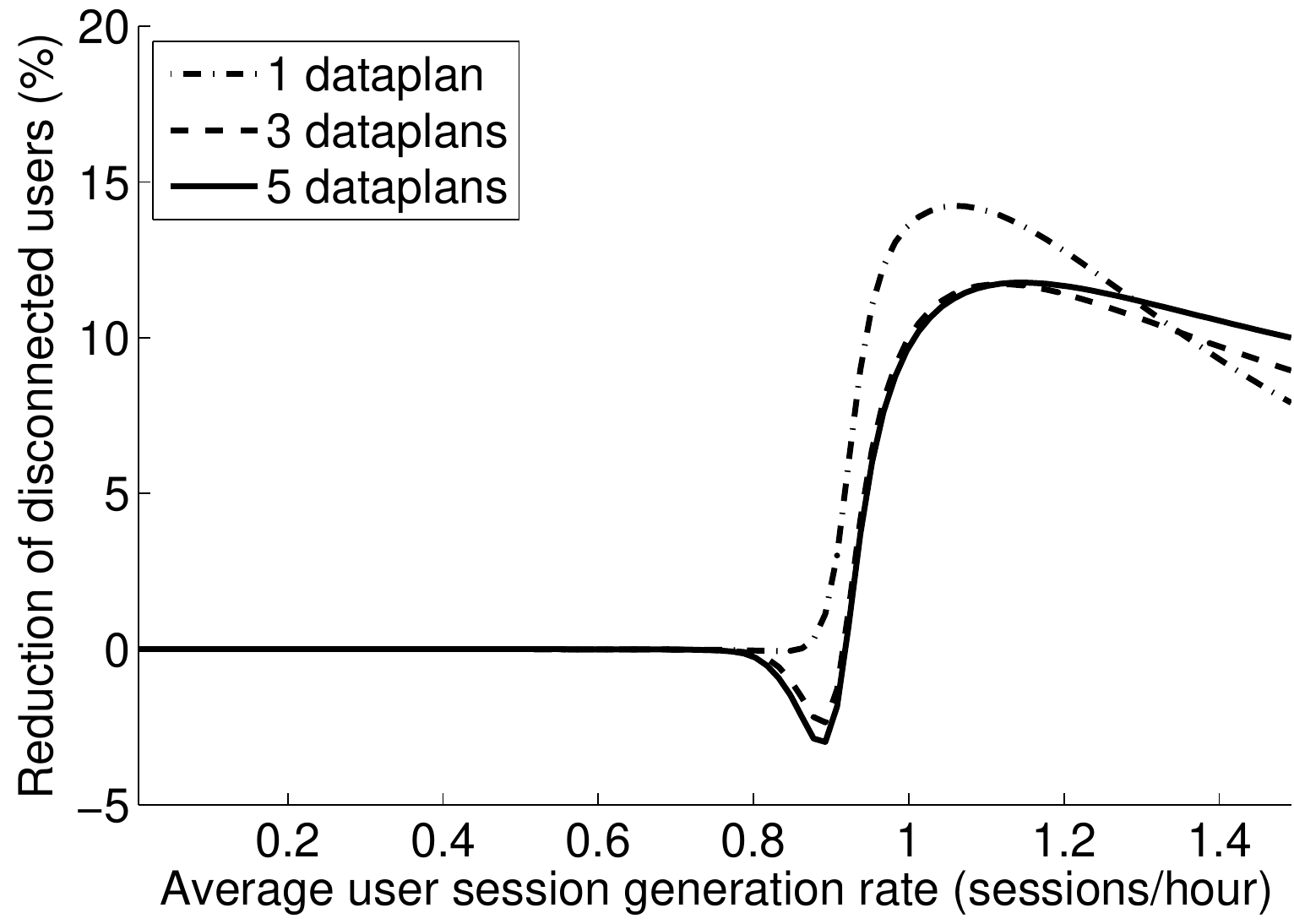}}
\subfloat[]{\label{dserv_corr_provider_1}\includegraphics[width=0.35\textwidth, height=42mm]{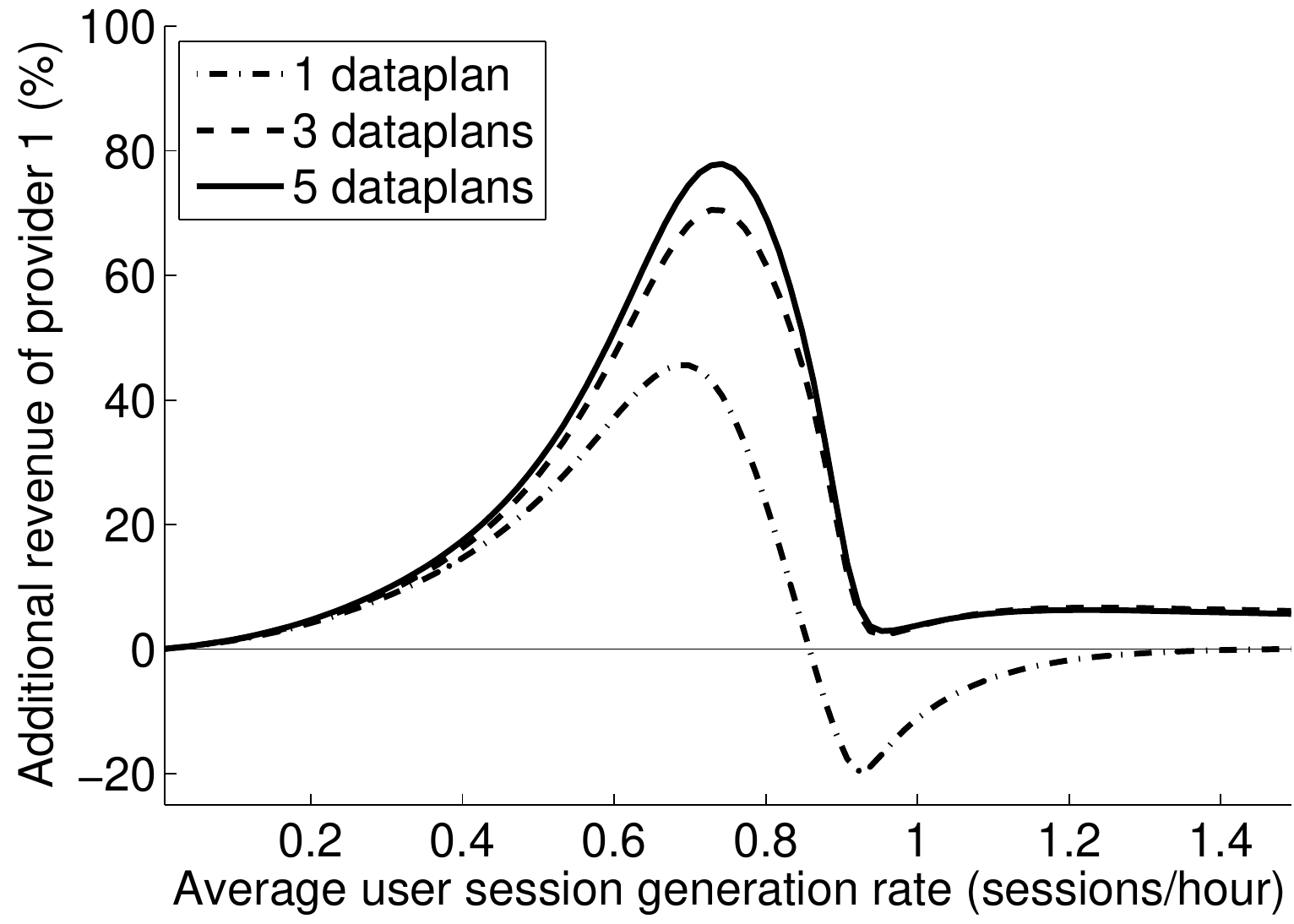}}
\subfloat[]{\label{dserv_corr_provider_4}\includegraphics[width=0.35\textwidth, height=42mm]{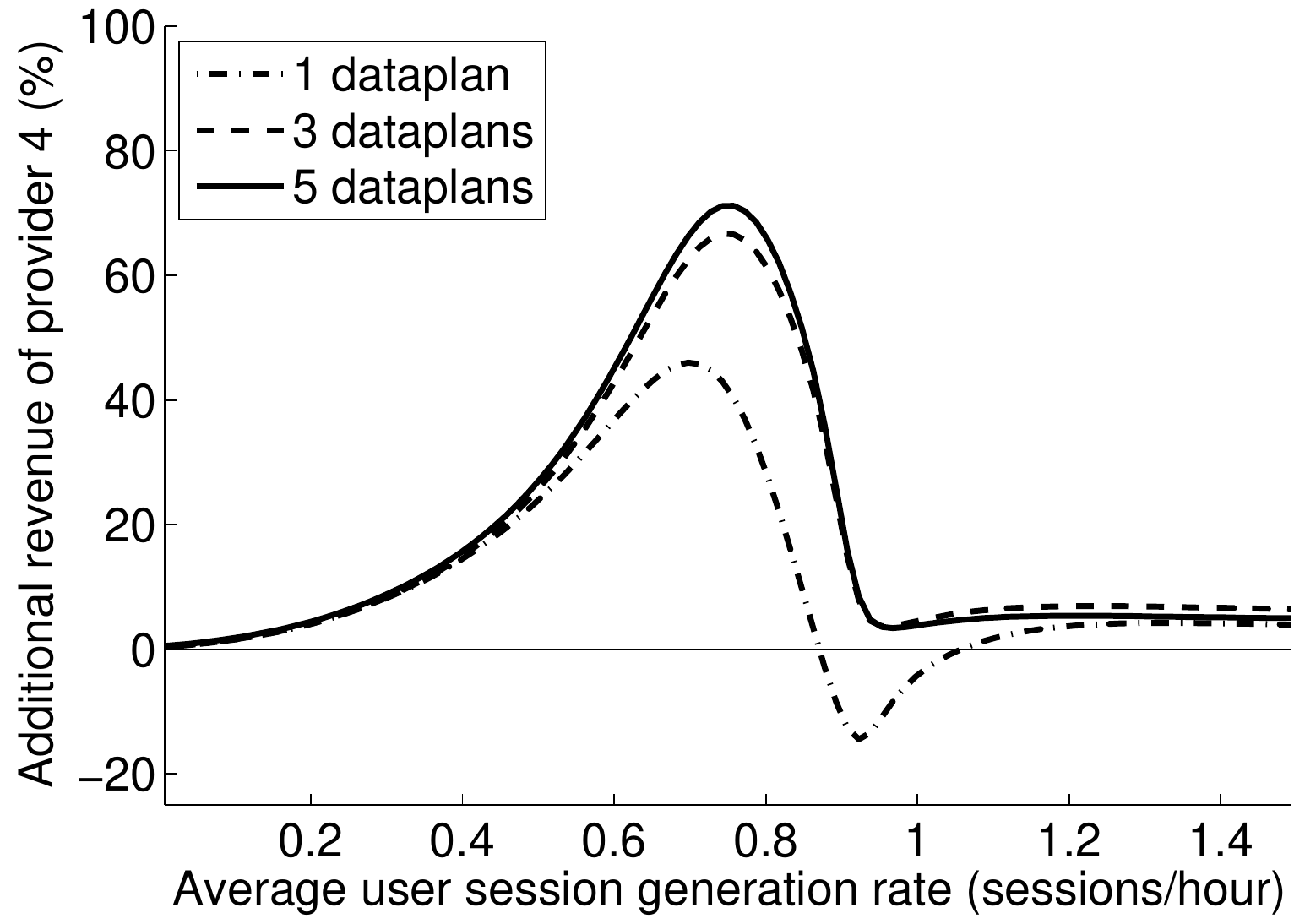}}\\
\subfloat[]{\label{dserv_ind_disc}\includegraphics[width=0.35\textwidth, height=42mm]{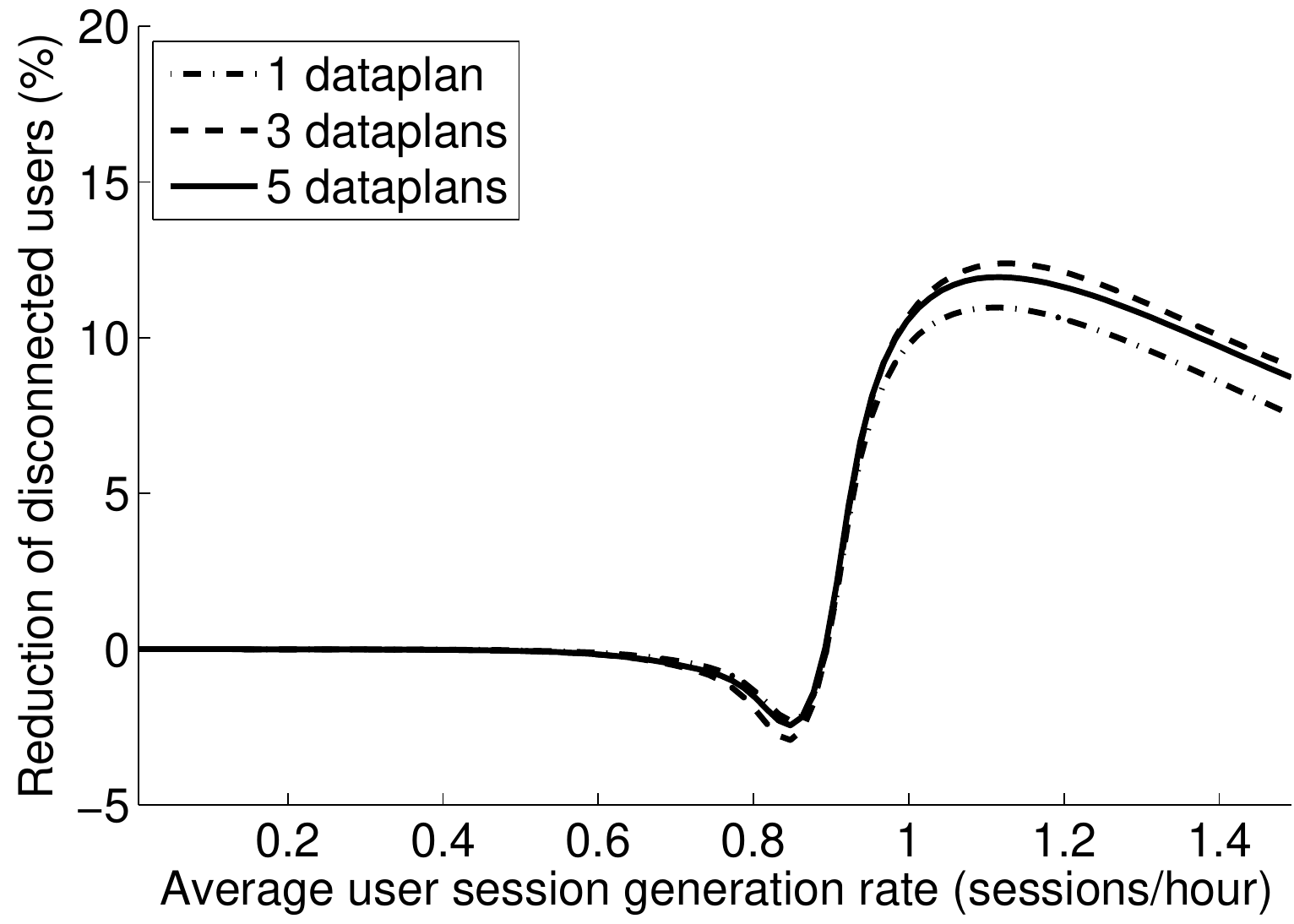}}
\subfloat[]{\label{dserv_ind_provider_1}\includegraphics[width=0.35\textwidth, height=42mm]{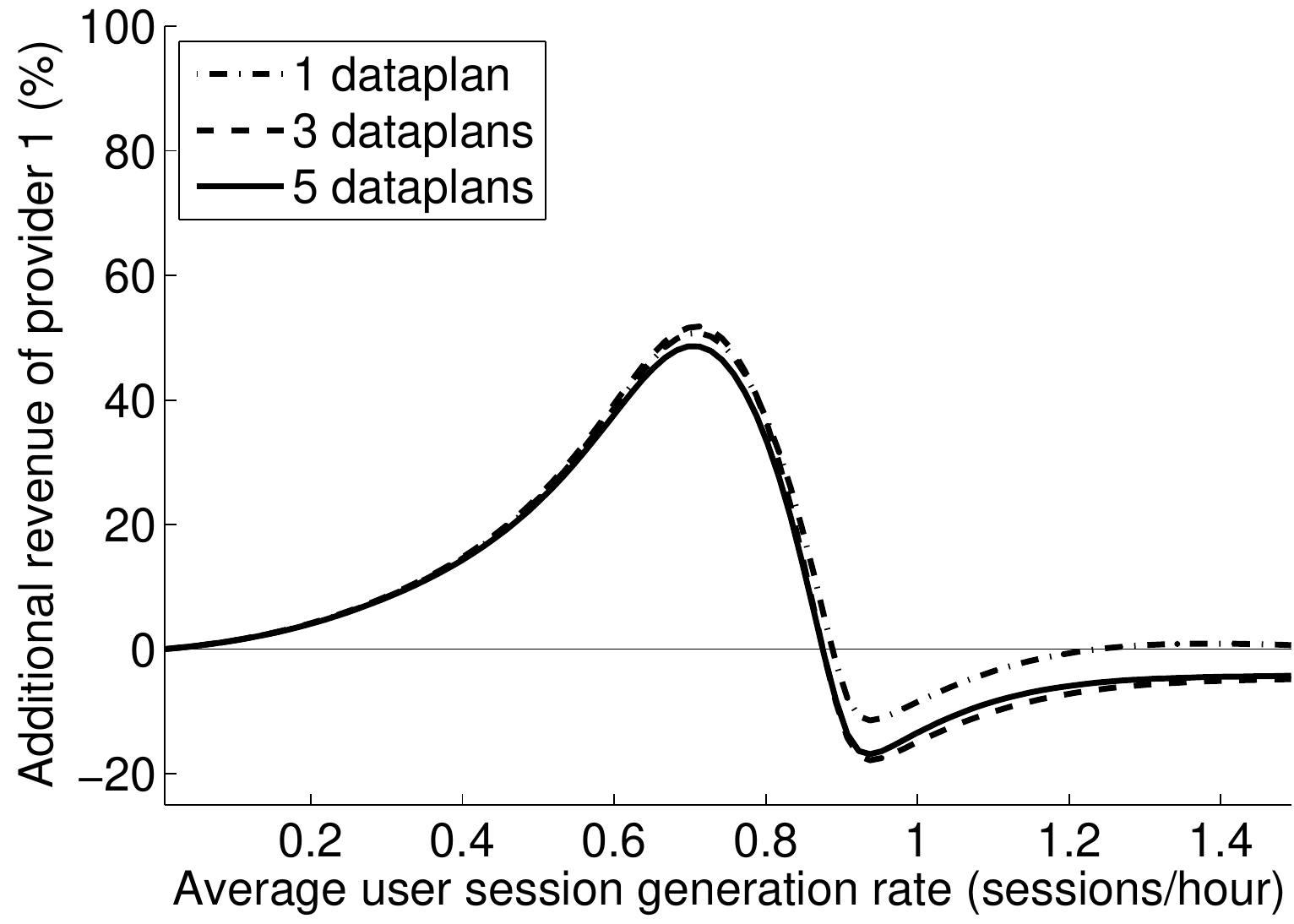}}
\subfloat[]{\label{dserv_ind_provider_4}\includegraphics[width=0.35\textwidth, height=42mm]{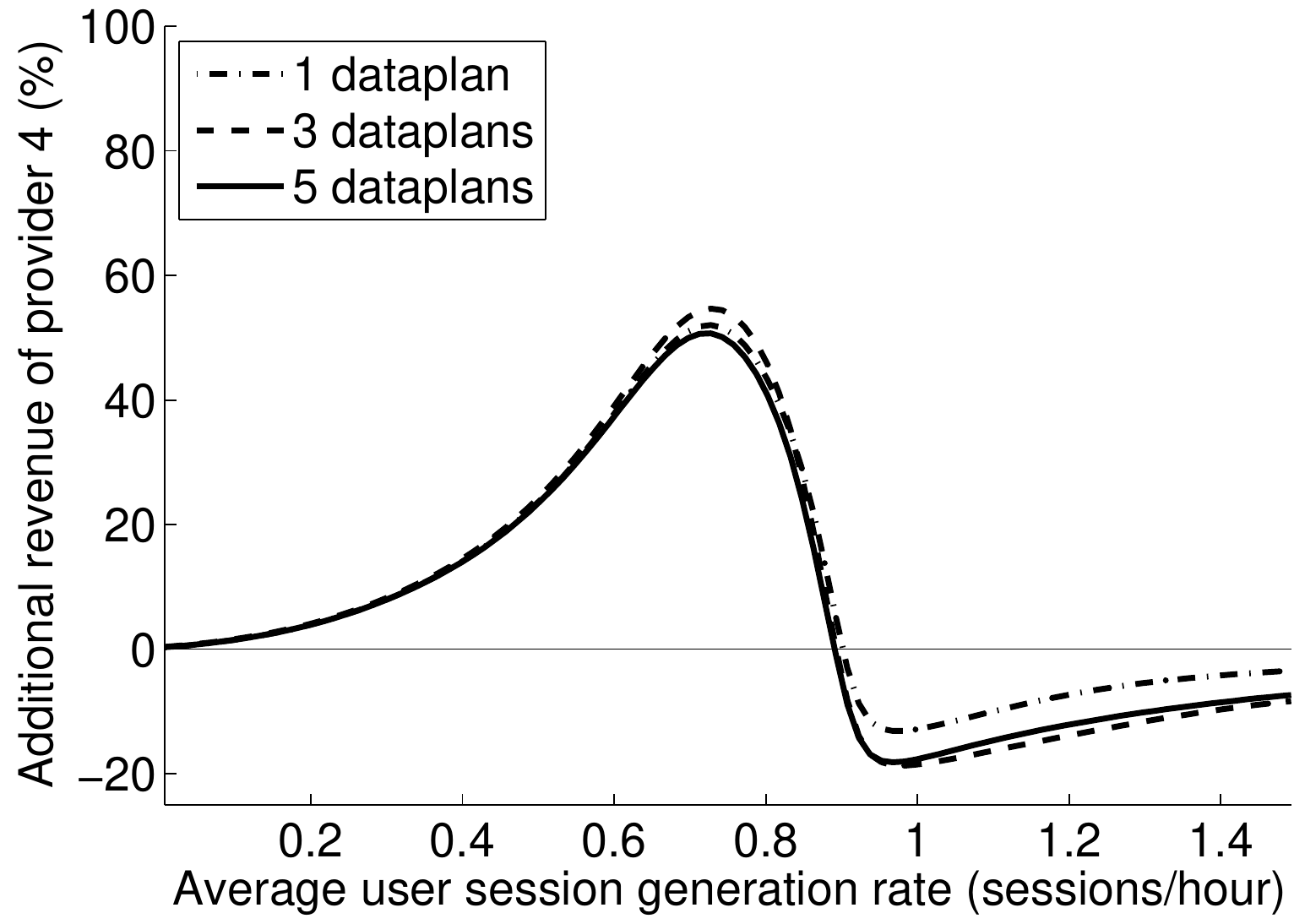}}
  \caption{Performance gains when each provider models users with $9$ clusters and offers a different number of dataplans compared to macroscopic modeling when $w_R^j$, $h_j$, and $n_j$ are correlated (top) and when they are independent (bottom), respectively.}
  \label{fig:dserv}
\end{figure*}

We have also studied a market in which providers offer a different number of dataplans (i.e., 1, 3, or 5 dataplans). In this market, each provider models the users with 9 clusters. The performance gains compared to a market in which providers model users macroscopically and offer only $1$ dataplan are depicted in Fig. \ref{fig:dserv}. In the top (bottom) figures the parameters $w^j_R$, $h_j$, and $n_j$ are correlated (independent), respectively. When $w^j_R$, $h_j$, and $n_j$ are correlated, if providers offer only $1$ dataplan, at an interval of the traffic demand around $0.9$ sessions/hour, they achieve revenue losses compared to the macroscopic case (Figs. \ref{dserv_corr_provider_1} and \ref{dserv_corr_provider_4}). Those losses are mainly due to the limited number of degrees of freedom when setting the prices of providers. Specifically, providers offer the same price both to heavy users with a large willingness to pay and to light users with a low willingness to pay. To prevent light users from becoming disconnected, providers restrict their prices regardless of the high willingness to pay of heavy users losing revenue. However, when providers offer $3$ or $5$ dataplans, the heavy and light users are charged with a different price. Therefore, providers are able to offer a high price to heavy users and a low price to light users always achieving revenue benefits (Figs. \ref{dserv_corr_provider_1} and \ref{dserv_corr_provider_4}). 

When $w^j_R$, $h_j$, and $n_j$ are independent, a counter intuitive result is observed. An increase in the number of offered dataplans does not result in revenue benefits for providers. On the contrary, under a large traffic demand, the offering of a larger number of dataplans results in revenue losses (Figs. \ref{dserv_ind_provider_1} and \ref{dserv_ind_provider_4}). In this case, the willingness to pay and traffic demand of user clusters are independent. This means that the average willingness to pay of heavy and light users is almost the same. When providers offer different dataplans to these users, their competition is intensified. Given that the light users produce a low amount of traffic, their admission at the network of a provider does not significantly affect its QoS. Therefore, providers have the incentive to reduce their offered prices to light users in order to attract them to their networks entering a price war. Additionally, providers can not charge the heavy users with a high price due to their relatively low willingness to pay (almost the same as the one of light users). Therefore, providers would lose revenue by offering a larger number of dataplans in this market. 
\section{Related work} \label{sec:Related_work}
The game-theoretical approaches in analysing wireless markets \cite{Tres12, Niyato08} can be classified into two general categories, namely, the microscopic approaches and the macroscopic ones. Microscopic models usually focus on a short spatial and temporal scale and evaluate the impact of various technical aspects (e.g., mobile data offloading \cite{Gao14, Zhuo14}, femtocells \cite{Lin14}, network selection mechanisms for users \cite{Gaji14, Anto07}, cooperative spectrum access schemes for primary and secondary users in cognitive radio networks \cite{Dape11, Wyso0410, Yang11}, and multihop access paradigms \cite{Yong11}) on the performance of a network, providers, and users satisfaction. These modeling approaches can be detailed and accurate but usually not scalable or amenable to theoretical analysis. On the other hand, macroscopic approaches focus on larger-scale phenomena and make various simplifications \cite{Jia09, Niya0710, Niya09, Niyat08, Jia0508}. For example, they may consider a homogeneous user population and model the aggregate user behaviour (e.g., traffic demand, QoS preferences \cite{Rose14, Ren11}), trading the accuracy for scalability and tractability. 
%
%
%

To the best of our knowledge, the proposed framework is one of the few game-theoretical approaches that models the users, network, and providers in detail. For example, it models relatively large-scale networks (at the BS level), the user demand, profile (e.g., sensitivity to the price and QoS, rationality and loyalty to certain providers), handovers, and mobility pattern.  
In our earlier work, we developed an event-based simulator of a wireless access duopoly that could be executed at multiple levels of detail from the microscopic to the macroscopic one \cite{forte0115}. However, no analytical results were derived for the user and provider equilibriums or the offered QoS and the generalization of the framework for a larger number of providers and dataplans was difficult. Our following work provided an analytical methodology for computing the Nash equilibriums (NEs) of users and providers in a wireless oligopoly at the macroscopic level \cite{forte0215}. Unfortunately its extension at higher levels of detail was challenging due to various technical difficulties. 

As mentioned in the introduction, this work extends our earlier papers and the state-of-the-art in the following ways: (1) It enables providers to model users at {\em different levels of detail} and {\em analytically} computes the equilibriums of users and providers. 
(2) It employs a detailed model of the network. 
(3) It allows providers to offer several dataplans depending on user traffic demand and models the user-decision making, capturing important aspects of customer behaviour.
(4) It relaxes the assumption about the user rationality in selecting providers and dataplans. 
\section{Conclusions and future work} \label{sec:Conclusion}

The proposed multi-layer two-stage game-theoretical modelling framework for wireless markets 
distinguishes different customer groups and models their decision making process. It also models providers with different capacities in their wireless infrastructure and degrees of knowledge about the customer population. The providers can offer multiple dataplans. 
To capture the user-decision making more realistically, the framework also models the ``stickiness'' to a provider. 

Under a macroscopic view of the market, providers make  suboptimal decisions. When providers model the customer population in a higher detail, they can improve their revenue. Gains also exist when providers offer a larger number of dataplans. 
In the case that the customer willingness to pay and tolerance on low data rate are independent, if providers share the same level of knowledge about users, their competition may become more severe, resulting in revenue loss. 
Additionally, the increase of the number of dataplans may also result in revenue losses due to the similar willingness to pay of heavy and light users in such scenario. 
When a provider models users at a high level of detail, its revenue strongly depends on the knowledge of the other providers about users. 
For example, if the other providers also model the users in a high degree of detail, the competition may get enhanced, resulting to revenue loss. 
On the other hand, when a provider models the customers at higher level of detail compared to the others, it can achieve significant benefits. The number of disconnected users also decreases. 
%

The analysis shows how providers develop their strategies and how they focus on specific customer segments.
For example, often the strong provider focuses on business-type of customers, while providers with less resources attract low-profile customers. However when the capacity of the network is reached, the strong provider may develop a different strategy to attract users with low willingness to pay.
Moreover offering a larger number of dataplans allows providers to charge the various customer segments more efficiently achieving a higher revenue, under the condition that the willingness-to-pay, tolerance in data rate and traffic are not independent. 

Service providers can apply such tools to assess the evolution of a market, for different customer profile densities, dataplans, and traffic demand. The number of providers and their capacity may also change to evaluate their impact on the market dynamics.
To address the congestion, cost reduction, and revenue generation, operators currently are also looking for attracting allies within the Internet market through sponsored data, offloading, infrastructure sharing, MVNOs, and network slicing. For example, in network slicing, an infrastructure provider offers network resources, service providers buy such resources according to the expected customer demand, and customers select the best plan/service provider. 
The proposed framework can be extended to analyze the dynamics and evolution of such new markets.

\bibliographystyle{plainnat}

\bibliography{references18}

\end{document}